\documentclass[a4paper,fleqn,usenatbib]{mnras}
\usepackage{natbib} 

\usepackage{newtxtext,newtxmath}

\usepackage[T1]{fontenc}
\usepackage{ae,aecompl}

\usepackage{graphicx}	
\usepackage{amsmath}	
\usepackage{amssymb}	

\newcommand{\cloudy}{\textsc{Cloudy}}
\newcommand{\eazy}{\textsc{EAZY}}
\newcommand{\sfrsed}{\ifmmode  \mathrm{SFR}_\mathrm{SED} \else $\mathrm{SFR}_\mathrm{SED}$\fi}
\newcommand{\sfruvir}{\ifmmode  \mathrm{SFR}_\mathrm{UV+IR} \else $\mathrm{SFR}_\mathrm{UV+IR}$\fi}
\newcommand{\sfruv}{\ifmmode  \mathrm{SFR}_\mathrm{UV} \else $\mathrm{SFR}_\mathrm{UV}$\fi}
\newcommand{\sfrlya}{\ifmmode  \mathrm{SFR}_{\mathrm{Ly}\alpha} \else $\mathrm{SFR}_{\mathrm{Ly}\alpha}$\fi}
\newcommand{\sfrha}{\ifmmode  \mathrm{SFR}_{\mathrm{H}\alpha} \else $\mathrm{SFR}_{\mathrm{H}\alpha}$\fi}
\newcommand{\flyc}{\ifmmode  \mathrm{f}_\mathrm{esc}\mathrm{(LyC)} \else $\mathrm{f}_\mathrm{esc}\mathrm{(LyC)}$\fi}
\newcommand{\flya}{\ifmmode  \mathrm{f}_\mathrm{esc}\mathrm{(Ly\alpha)} \else $\mathrm{f}_\mathrm{esc}\mathrm{(Ly}\alpha)$\fi}
\newcommand{\fefflya}{\ifmmode  \mathrm{f}_\mathrm{esc}^\mathrm{eff}\mathrm{(Ly\alpha)} \else $\mathrm{f}_\mathrm{esc}^\mathrm{eff}\mathrm{(Ly}\alpha)$\fi}
\newcommand{\flyarel}{\ifmmode  \mathrm{f}_\mathrm{esc}^\mathrm{rel}\mathrm{(Ly\alpha)} \else $\mathrm{f}_\mathrm{esc}^\mathrm{rel}\mathrm{(Ly}\alpha)$\fi}
\newcommand{\dfel}{\ifmmode \Delta\log {\rm f}_{\rm EL} \else $\Delta\log$ f$_{\rm EL}$\fi}
\newcommand{\hst}{{\it HST}}
\newcommand{\jwst}{{\it JWST}}
\newcommand{\spitzer}{{\it Spitzer}}
\newcommand{\hyperz}{{\it Hyperz}}
\newcommand{\flyased}{\ifmmode  \mathrm{f}_\mathrm{SED}\mathrm{(Ly\alpha)} \else $\mathrm{f}_\mathrm{SED}\mathrm{(Ly}\alpha)$\fi}
\newcommand{\flyaobs}{\ifmmode  \mathrm{f}_\mathrm{obs}\mathrm{(Ly\alpha)} \else $\mathrm{f}_\mathrm{obs}\mathrm{(Ly}\alpha)$\fi}
\newcommand{\tigm}{\ifmmode  \mathrm{T}_\mathrm{IGM} \else $\mathrm{T}_\mathrm{IGM}$\fi}
\newcommand{\nh}{$n$(\textrm{H})}
\newcommand{\CO}{$\log$(\textrm{C}/\textrm{O})}
\newcommand{\U}{\ifmmode  \log\mathrm{U}  \else $\log\mathrm{U}$ \fi}

\newcommand{\oiii}{[\textrm{O}~\textsc{iii}]}
\newcommand{\oii}{[\textrm{O}~\textsc{ii}]}
\newcommand{\oiilam}{[\textrm{O}~\textsc{ii}]\ensuremath{\lambda3727}}
\newcommand{\oiiiv}{[\textrm{O}~\textsc{iii}]\ensuremath{\lambda5007}}
\newcommand{\oiiiiv}{[\textrm{O}~\textsc{iii}]\ensuremath{\lambda4959}}

\newcommand{\oiiidoub}{[\textrm{O}~\textsc{iii}]\ensuremath{\lambda\lambda4959,5007}}
 
\newcommand{\ha}{\ifmmode {\rm H}\alpha \else H$\alpha$\fi}
\newcommand{\hb}{\ifmmode {\rm H}\beta \else H$\beta$\fi}
\newcommand{\lya}{\ifmmode {\rm Ly}\alpha \else Ly$\alpha$\fi}
\newcommand{\pg}{\ifmmode {\rm P}\gamma \else P$\gamma$\fi}
\newcommand{\lyb}{\ifmmode {\rm Ly}\beta \else Ly$\beta$\fi}
\newcommand{\lyg}{\ifmmode {\rm Ly}\gamma \else Ly$\gamma$\fi}

\newcommand{\ciiidoub}{\textrm{C}~\textsc{iii}]\ensuremath{\lambda\lambda1907,1909}}

\newcommand{\civdoub}{\textrm{C}~\textsc{iv}\ensuremath{\lambda\lambda 1548,1550}}

\newcommand{\msun}{\ifmmode \mathrm{M}_{\odot} \else M$_{\odot}$\fi}
\newcommand{\msunyr}{\ifmmode \mathrm{M}_{\odot} {\rm yr}^{-1} \else $\mathrm{M}_{\odot}$ yr$^{-1}$\fi}
\newcommand{\zsun}{\ifmmode Z_{\odot} \else Z$_{\odot}$\fi}
\newcommand{\lsun}{\ifmmode L_{\odot} \else L$_{\odot}$\fi}
\newcommand{\mstar}{\ifmmode \mathrm{M}_{\star} \else M$_{\star}$\fi}

\newcommand{\myemail}{stephane.debarros@unige.ch}
\newcommand{\myinstitute}{D\'{e}partement d'Astronomie, Universit\'{e} de Gen\`{e}ve, 51 Ch. des Maillettes, 1290 Versoix, Switzerland}

\title[The GREATS H$\beta$+{[O\,{\normalsize \textrm{III}}]} Luminosity Function at $z\sim8$]{The GREATS H$\beta$+{[O\,{\normalsize \textrm{III}}]} Luminosity Function and Galaxy Properties at $\mathbf{z\sim8}$: Walking the Way of \jwst}

\author[S.~De Barros et al.]{S.~De Barros$^{1}$\thanks{E-mail: \myemail},
P.~A.~Oesch$^{1,2}$, I. Labb\'{e}$^{3}$,
M.~Stefanon$^{4}$,
V.~Gonz\'{a}lez$^{5,6}$,\newauthor
R.~Smit$^{7,8}$,
R.~J.~Bouwens$^{4}$,
G. D. Illingworth$^{9}$
\\
$^{1}$\myinstitute\\
$^{2}$International Associate, Cosmic Dawn Center (DAWN) at the Niels Bohr Institute, University of Copenhagen and DTU-Space,\\ Technical University of Denmark\\
$^{3}$Centre for Astrophysics \& Supercomputing, Swinburne University of Technology, PO Box 218, Hawthorn, VIC 3112, Australia\\
$^{4}$Leiden Observatory, Leiden University, NL-2300 RA Leiden, Netherlands\\
$^{5}$Departmento de Astronomia, Universidad de Chile, Casilla 36-D, Santiago 7591245, Chile\\
$^{6}$Centro de Astrofisica y Tecnologias Afines (CATA), Camino del Observatorio 1515, Las Condes, Santiago 7591245, Chile\\
$^{7}$Cavendish Laboratory, University of Cambridge, 19 JJ Thomson Avenue, Cambridge CB3 0HE, UK\\
$^{8}$Kavli Institute for Cosmology, University of Cambridge, Madingley Road, Cambridge CB3 0HA, UK \\
$^{9}$UCO/Lick Observatory, University of California, Santa Cruz, CA 95064, USA\\
}

\date{Accepted XXX. Received YYY; in original form ZZZ}

\pubyear{2018}

\begin{document}
\label{firstpage}
\pagerange{\pageref{firstpage}--\pageref{lastpage}}
\maketitle

\begin{abstract}
The {\it James Webb Space Telescope} will allow to spectroscopically study an unprecedented number of galaxies deep into the reionization era, notably by detecting \oiiidoub\ and \hb\ nebular emission lines.
To efficiently prepare such observations, we photometrically select a large sample of galaxies at $z\sim8$ and study their rest-frame optical emission lines.
Combining data from the GOODS Re-ionization Era wide-Area Treasury from \spitzer\ (GREATS) survey and from \hst\, we perform spectral energy distribution (SED) fitting, using synthetic SEDs from a large grid of photoionization models.
The deep \spitzer/IRAC data combined with our models exploring a large parameter space enables to constrain the \oiii+\hb\ fluxes and equivalent widths for our sample, as well as the average physical properties of $z\sim8$ galaxies, such as the ionizing photon production efficiency with $\log(\xi_\mathrm{ion}/\mathrm{erg}^{-1}\hspace{1mm}\mathrm{Hz})\geq25.77$.
We find a relatively tight correlation between the \oiii+\hb\ and UV luminosity, which we use to derive for the first time the \oiiidoub+\hb\ luminosity function (LF) at $z\sim8$.
The $z\sim8$ \oiii+\hb\ LF is higher at all luminosities compared to lower redshift, as opposed to the UV LF, due to an increase of the \oiii+\hb\ luminosity at a given UV luminosity from $z\sim3$ to $z\sim8$.
Finally, using the \oiii+\hb\ LF, we make predictions for \jwst/NIRSpec number counts of $z\sim8$ galaxies. We find that the current wide-area extragalactic legacy fields are too shallow to use \jwst\ at maximal efficiency for $z\sim8$ spectroscopy even at 1hr depth and \jwst\ pre-imaging to $\gtrsim30$ mag will be required.

\end{abstract}

\begin{keywords}
galaxies: evolution -- galaxies: high-redshift -- reionization
\end{keywords}



\section{Introduction}
\label{sec:intro}

Great progress has been made over the last two decades in our study of early galaxy mass assembly and the evolution of the cosmic star-formation rate density at $z\geq4$ \citep[e.g.,][]{madaudickinson14,duncan+14,salmon+15}. However, so far, these studies mostly rely on the analysis of broad-band photometry only, given that current facilities only provide access to the faint, rest-UV emission lines in a small number of bright galaxies \citep[e.g.,][]{stark+17}.
While a wealth of photometric data are now publicly available, photometric studies can suffer from several caveats. The selection of high-redshift galaxies relies on the Lyman break \citep[i.e., dropout selection;][]{steidel+96} that can possibly miss a significant fraction of galaxies \citep[e.g.,][]{inami+17}, and the derivation of most of the galaxy physical properties relies on spectral energy distribution (SED) fitting that is affected by several degeneracies and strongly depends on assumptions \citep[e.g., star formation history, metallicity, dust extinction curve;][]{finlator+07,yabe+09,debarros+14}. Furthermore, the photometry can be contaminated by strong nebular emission lines that are ubiquitous at high-redshift \citep[e.g.,][]{chary+05,schaererdebarros10,shim+11,stark+13,labbe+13,smit+14,shivaei+15a,faisst+16,marmol+16,rasappu+16}. To account for their impact, either empirical \citep[e.g.,][]{schaererdebarros09} or dedicated photoionization modeling \citep[e.g.,][]{zackrisson+01} have been used. However, direct observational access to these emission lines through spectroscopy will have to await the advent of the {\it James Webb Space Telescope} (\jwst).

	At lower redshift, where most ultraviolet, optical, and near-infrared emission lines can be observed either from the ground or space, lines are efficiently used to determine instantaneous star-formation rates (SFR) and specific star-formation rates 
	\citep[sSFR=SFR/\mstar;][]{kauffmann+04}, to derive gas-phase element abundances \citep[e.g.,][]{tremonti+04}, to accurately derive the dust extinction thanks to the Balmer decrement \citep[e.g.,][]{dominguez+13,reddy+15}, and to determine the main source of ionizing photons (star formation or AGN) with the BPT diagram \citep{baldwin+81}. While direct observations of optical and near-infrared lines are out of reach for $z>4$ galaxies until the launch of \jwst, one can take advantage of the impact of nebular emission lines on the photometry to probe these lines indirectly and to uniquely reveal some ISM properties of high-redshift galaxies.
	
	Emission lines can also be useful to derive very accurate photometric redshifts. 
	\cite{smit+15} exploit extremely blue \spitzer/IRAC colors to identify $6.6\leq z\leq6.9$ galaxies, for which \oiiidoub+\hb\ lines are expected to fall in the 3.6\micron\ band while the 4.5\micron\ band is free of line contamination.
	A similar technique applied to $z>7.1$ galaxies has led to the reliable selection and subsequent spectroscopic confirmation of some of the most distant Lyman-$\alpha$\ emitters to date \citep{robertsborsani+16,oesch+15,zitrin+15,stark+17}. SFR and sSFR can be derived from \ha\ emission for galaxies at $z\sim4$ where the \ha\ line is found in the IRAC 3.6\micron\ channel while the IRAC 4.5\micron\ channel is free from line contamination \citep{shim+11,stark+13,marmol+16}. Furthermore, the \ha\ luminosities have also been used to derive the  ionizing photon production efficiency \citep[$\xi_\mathrm{ion}$, defined as the production rate of ionizing photons per unit luminosity in the UV-continuum;][]{bouwens+16a} at $z\sim4$.  The comparison between uncorrected SFR derived from emission lines and SFR derived from UV+IR shed light on the relative stellar to nebular attenuation \citep{shivaei+15a,debarros+16b}.
	
	In this paper, we adopt the same approach as these latter studies: we use the impact of nebular emission on the broad-band photometry to indirectly 
	derive emission line fluxes and EWs to gain insight into high-redshift galaxy properties. Specifically, we use a sample of photometrically selected $z\sim8$ galaxies, for which the \oiiidoub+\hb\ lines fall in the IRAC 4.5\micron\ channel while we do not expect strong lines in the IRAC 3.6\micron\ channel. We take advantage of the \spitzer\ ultra-deep survey covering CANDELS/GOODS South and North fields, the GOODS Re-ionization Era wide-Area Treasury from Spitzer (GREATS, Labb\'{e} et al. 2019, in prep) survey, providing the best constraints on $z\sim8$ IRAC colors to date.	To derive  the line fluxes as accurately as possible and account for most of the uncertainties, we use synthetic SEDs produced with dedicated photoionization modeling to fit the observed $z\sim8$ SEDs. Our aim is to derive the \oiiidoub+\hb\ luminosity function (LF) at $z\sim8$ to prepare efficient \jwst\ observations in the future.
	
	The paper is structured as follows: The photometric data and selection procedure are described in Sec.~\ref{sec:data}. We provide a description of our photoionization grid in Sec.~\ref{sec:phot}, and Sec.~\ref{sec:sed} gives the SED fitting method. In Sec.~\ref{sec:ism}, we present the constraints that we obtain on physical properties of $z\sim8$ galaxies. The resulting \oiiidoub+\hb\ luminosity function is shown and discussed in Sec.~\ref{sec:LF}. We summarize our conclusions in Sec.~\ref{sec:conc}.

We adopt a $\Lambda$-CDM
cosmological model with $\mathrm{H}_0=70$ km s$^{-1}$ Mpc$^{-1}$,
$\Omega_{\mathrm{m}}=0.3$ and $\Omega_{\Lambda}=0.7$. All magnitudes are expressed in the
AB system \citep{okegunn83}.

\section{Data and Sample}
\label{sec:data}

The input sample used for this work is based on the Lyman break galaxy (LBG) catalogs from \citet{bouwens+15b}. These are compiled from all the prime extragalactic legacy fields, including the Hubble Ultra Deep Field \citep[HUDF;][]{ellis+13,illingworth+13} and its parallel fields, as well as all five CANDELS fields \citep{grogin+11,koekemoer+11}. 

In addition to deep $HST$ near-infrared imaging, all these fields have extensive $Spitzer$/IRAC coverage. We have reduced and combined all the IRAC 3.6 and 4.5 $\micron$ data available in each field. In particular, we include the complete data from the GREATS survey (Labbe et al. 2019, in prep). GREATS builds on the vast amount of archival data in the two GOODS fields \citep{giavalisco+04a} and brings the IRAC 3.6\micron\ and 4.5\micron\ coverage to a near-homogeneous depth of 200-250 hr, corresponding to 5$\sigma$ sensitivities of 26.8-27.1 mag, over $\sim200$ arcmin$^2$. This is very well matched to the $H_{160}$-band detection limits, allowing us to detect the rest-frame optical light of nearly all the galaxies identified in the $HST$ data. For a complete description of the $HST$ and $Spitzer$ dataset we refer the reader to \citet{bouwens+15b} and Stefanon et al. (2019, in prep).

Given the much wider point-spread function (PSF) of the IRAC data compared to $HST$, special care is required to derive accurate photometry. We use a custom-made software tool \texttt{mophongo}, developed and updated over the last few years \citep[e.g.,][]{labbe+10,labbe+15}. In short, starting from the $HST$ F160W image, \texttt{mophongo} uses position-dependent $HST$-to-IRAC PSF kernels to fit and subtract all the neighboring galaxies in a 21\arcsec\ region around a source of interest, before measuring its flux density in a 2\arcsec\ aperture.

   \begin{figure}
          \centering
     \includegraphics[width=\columnwidth,trim=1cm 0cm 1.5cm 1cm,clip=true]{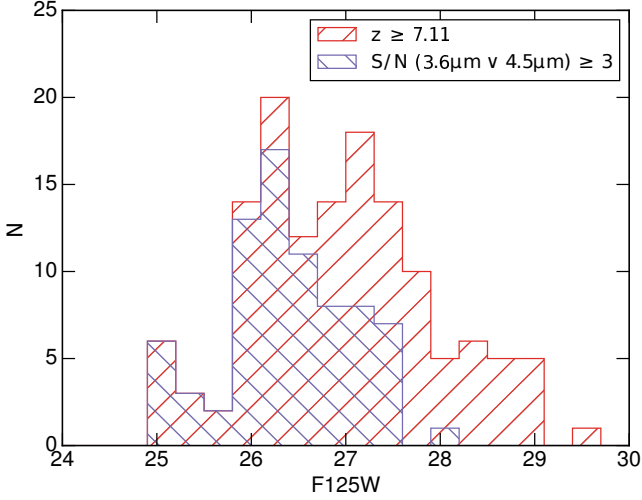}
     \caption{F125W magnitude distribution for the sample before (red) and after applying the 3.6\micron\ and/or 4.5\micron\ detection ($S/N>3$) requirement (purple). This latter criterion favors UV bright galaxies with F125W$<27.5$.}
     \label{fig:fig1}
 \end{figure} 

After discarding objects with IRAC fluxes highly contaminated by neighbours ($<10\%$), we ran the \eazy\ code \citep{brammer+08} on the full photometric catalog, allowing the photometric redshift to vary from 0 to 10, and deriving the redshift probability distribution for each object in the GREATS data. We took care to exclude IRAC 3.6\micron\ and IRAC 4.5\micron\ bands from the SED fit, since emission lines at high redshift can affect the photometry in these bands \citep[e.g.,][]{smit+14} and we want to avoid to be biased toward objects with large  EW(\oiiidoub+\hb). We want to focus on $z\sim8$ galaxies for which the IRAC 3.6\micron$-$IRAC 4.5\micron\ color provides constraints on the \oiiidoub+\hb\ flux and these lines have completely entered the \spitzer/IRAC 4.5\micron\ channel at $z\geq7.11$ \footnote{The \oiii+\hb\ lines are out of IRAC 4.5\micron\ at $z>9.05$ and after applying our selection, only one galaxy has a redshift above this limit.}. Therefore, the selection criterion was defined as $p(z\geq7.11)\geq0.68$, with $p(z\geq7.11)$ the probability for a galaxy to have a redshift $z\geq7.11$. Additionally to this criterion, we select a subsample of galaxies with at least one detection with $S/N\geq3$ in either IRAC 3.6\micron\ or IRAC 4.5\micron\, allowing us to derive at least a strong upper or lower limit on the $(3.6-4.5)$\micron\ color for each of these galaxies. In this subsample ($N=76$), 16 galaxies are detected at 4.5\micron\ only and 10 at 3.6\micron\ only, potentially slightly biasing this subsample toward larger EW(\oiiidoub+\hb), with the detection at 4.5\micron\ being consistent with an increase in flux due to \oiii+\hb\ lines, while the stellar continuum could remain undetected at 3.6\micron.  We show in Fig.~\ref{fig:fig1} the F125W magnitude distribution for the galaxy selected based on their photometric redshifts ($N=135$) and the same distribution after applying the {\it Spitzer}/IRAC detection criterion. The F125W band probes the UV rest-frame emission of $z\sim8$ galaxies ($\sim1500$\AA). $\sim50\%$ of the $z\sim8$ galaxies are detected with $S/N\geq3$ in at least one {\it Spitzer}/IRAC band, mostly the brightest with F125W$<27.5$. While we apply a threshold detection in IRAC to select our subsample, we use all available photometry including bands with low S/N (<3) as well as non detections to perform the SED fitting. We use the subsample ($N=76$) with $S/N(3.6\mu\mathrm{m}\lor4.5\mu\mathrm{m})\geq3$ to constrain the $z\sim8$ galaxy physical properties (Sec.~\ref{sec:ism}) and the entire photometric sample ($N=135$) to derive the \oiii+\hb\ LF (Sec.~\ref{sec:LF}).

\section{Photoionization models}
\label{sec:phot}

Several works have used photoionization models to study or predict nebular emission line properties of high-redshift galaxies \citep[e.g.,][]{zackrisson+11,jaskotravindranath16,steidel+16,nakajima+18,berg+18}.
Since we do not have access with the current facilities to the optical/NIR nebular emission lines for galaxies at $z\sim8$, the ISM physical conditions at these high redshifts are largely unknown. Therefore we created a grid of photoionization models with a large parameter space to encompass the plausible stellar and ISM physical properties at $z\sim8$

We used the latest release of the \cloudy\ photoionization code \citep[C17,][]{ferland+17} to build our grid of models. We chose SEDs from the latest 
BPASSv2.1 models \citep{eldridge+17} as input, which account for stellar binaries and stellar rotation effects. Indeed, several recent observations point out that high-redshift galaxies can exhibit UV emission lines requiring hard ionizing photons \citep[e.g., \ciiidoub, \civdoub;][]{stark+14,amorin+17,vanzella+17}, and there are mounting evidences that these UV lines are due to star formation since they are spatially associated with star forming regions \citep{smit+17}. These kind of strong emission lines can be reproduced by including stellar rotation and/or stellar binaries in the stellar models \citep[e.g.,][]{elrdidge+08} or using stellar templates updated with recent UV spectral libraries and stellar evolutionary tracks as in the latest Charlot \&\ Bruzual single star models \citep[e.g.,][]{gutkin+16}. Regarding the ISM properties, we adopt the same interstellar abundances and depletion factors of metals on to dust grains, and dust properties as \cite{gutkin+16}. These authors show that these modeling assumptions span a range that can reproduce most of the observed UV and optical emission lines at low- and high-redshift \citep{stark+14,stark+15a,stark+15b,stark+17,chevallard+16}.
While \cite{gutkin+16} use different stellar population synthesis model than used here, namely  an updated version of the \cite{BC03} stellar population synthesis model, a comparison of these two SPS models show that they provide similar results in interpreting stellar and nebular emissions of local massive star-clusters \citep{wofford+16}. For the grid used in this work, we use stellar metallicities from $Z=0.001$ to $Z=0.008$ with an initial mass function (IMF) index of $-2.35$ and an upper mass cutoff of 300\msun. For each stellar metallicity, for simplicity, we assume the same gas-phase metallicity.
We also explore a range of C/O abundance ratio \citep[from $\log\mathrm{C}/\mathrm{O}=-1.0$ to -0.4, consistent with the observations of][]{amorin+17}, three different values of dust-to-metal ratios \citep[$\xi_d=0.1,0.3,0.5$;][]{gutkin+16}, and a range of hydrogen gas densities ($10^2$ to $10^3$ cm$^{-3}$). We assume no leakage of ionizing photons. For each set of parameters and each stellar age, we built SEDs, adding to the pure stellar SEDs from BPASS the nebular emission lines and nebular continuum as computed in \cloudy.

\section{SED fitting}
\label{sec:sed}

We use the spectral energy distributions created with \cloudy\ to perform SED fitting of each individual galaxy in our $z\sim8$ sample. We use a modified version of the SED fitting code \hyperz\ \citep{bolzonella+00} allowing us to apply two different attenuation curves to the stellar and nebular components of the SED. We apply a Calzetti attenuation curve \citep{calzetti+00} to the stellar component and a Cardelli attenuation curve \citep{cardelli+89} to the nebular component, adopting $E(B-V)_\mathrm{gas}=E(B-V)_\mathrm{\star}$ for simplicity. Studies of galaxy samples at $z\sim2$ show that the ratio $E(B-V)_\mathrm{gas}/E(B-V)_\mathrm{\star}$ is affected by a large scatter and is increasing with increasing $E(B-V)_\mathrm{\star}$ and SFR \citep{reddy+15,theios+19}. Since $z\sim8$ galaxies exhibit blue UV $\beta$ slopes indicating low dust extinction \citep[e.g.,][]{bouwens+14}, we do not expect dust to have a large impact on our results. Nevertheless we allow $E(B-V)$ to vary from 0.0 to 0.2 in our SED fitting procedure. We assume a constant star formation history. While the choice of the SFH has an impact on the derived physical parameters, this effect is alleviated for young ages \citep[$<100$Myr;][]{debarros+14}. Since the oldest age allowed by the cosmological model adopted in this work is 730 Myr and due to the relatively young best-fit ages found for our sample (Sec.~\ref{sec:ism}), the assumed SFH has a limited impact on our final results.

Minimization of $\chi^2$ over the entire parameter space yields the best-fit SED. For each physical parameter of interest, we derive the median of the marginalized likelihood, and its associated uncertainties.
Derived physical parameters include age of the stellar population, stellar mass, SFR, sSFR, as well as observed (i.e., attenuated by dust) \oiiidoub\ and \hb\ emission line fluxes and EWs, and ISM physical properties (e.g., ionization parameter). All physical properties used in this work such as EW(\oiii+\hb) and L(\oiii+\hb) are SED derived, and correspond to the median of the marginalized likelihood, except stated otherwise.

The median uncertainty regarding the EW(\oiiidoub+\hb) for the entire sample is $+0.34/-0.37$ dex. We also split the sample in four bins of F125W magnitude to emphasize the reliability of the constraints depending on the UV luminosity. For $\mathrm{F125W}<26$,  $26<\mathrm{F125W}<26.5$, $26.5<\mathrm{F125W}<27$, and $\mathrm{F125W}>27$, the median uncertainties are $+0.27/-0.32$ dex, $+0.30/-0.32$ dex, $+0.38/-0.45$ dex, and $+0.39/-0.52$ dex, respectively. We describe how we account for those relatively large uncertainties in the \oiii+\hb\ luminosity function (LF) derivation in Sec.~\ref{sec:uvoiii}.

 \begin{figure*}
      \centering
     \includegraphics[width=\textwidth,trim=1.5cm 0cm 3.5cm 0cm,clip=true]{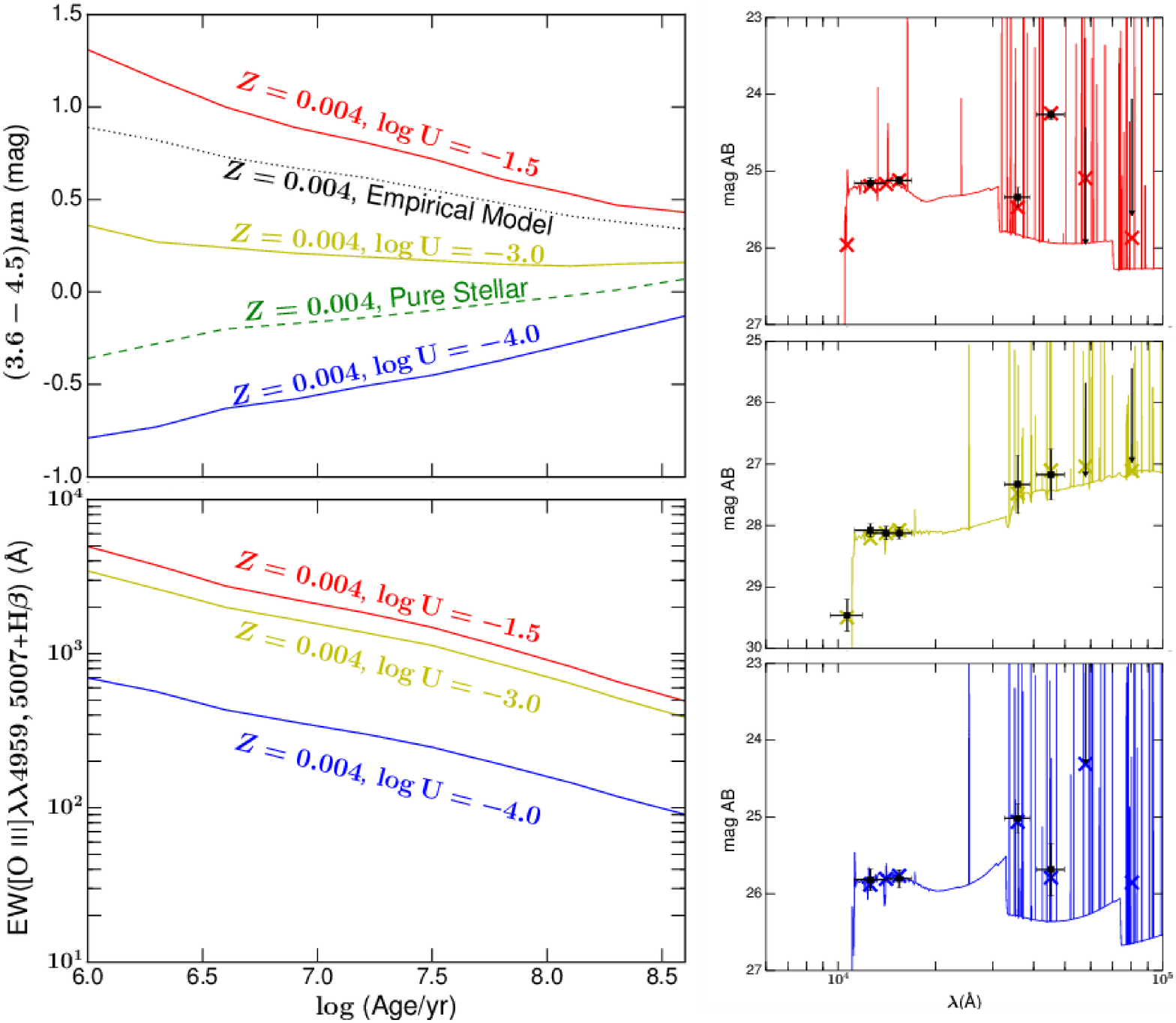}
     \caption{{\bf Top and bottom left panels:} Range of IRAC 3.6\micron$-$IRAC 4.5\micron\ colors (at $z=7.5$) and EW(\oiiidoub+\hb) vs. age probed by our grid of photoionization models for $Z=0.004$ and $\log\mathrm{U}=-1.5$, $-3.0$, and $-4.0$ in red, yellow, and blue, respectively.  The hydrogen densities \nh\ of the models shown lie between $10^2$ and $10^3$ (the impact of a \nh\ variation in this range is small) and we do not specify the carbon to oxygen abundance \CO\ since this parameter also has no impact on the quantities shown. Also shown are IRAC colors for a pure stellar  BPASSv2.1 template (dashed green line) and a template using typical empirical modeling of nebular emission \citep[continuum and lines;][]{schaererdebarros09,schaererdebarros10}. {\bf Right panels:} Examples of three best-fit SEDs obtained with the models shown on the left panels (same colors) for $z\sim7.5-8$ galaxies from our broader LBG sample. Clearly, the IRAC colors are heavily affected by the vast amount of emission lines. }
\label{fig:fig2}
 \end{figure*} 
 
   \begin{figure*}
      \centering
     \includegraphics[width=\textwidth,trim=0cm 14.5cm 0cm 0cm,clip=true]{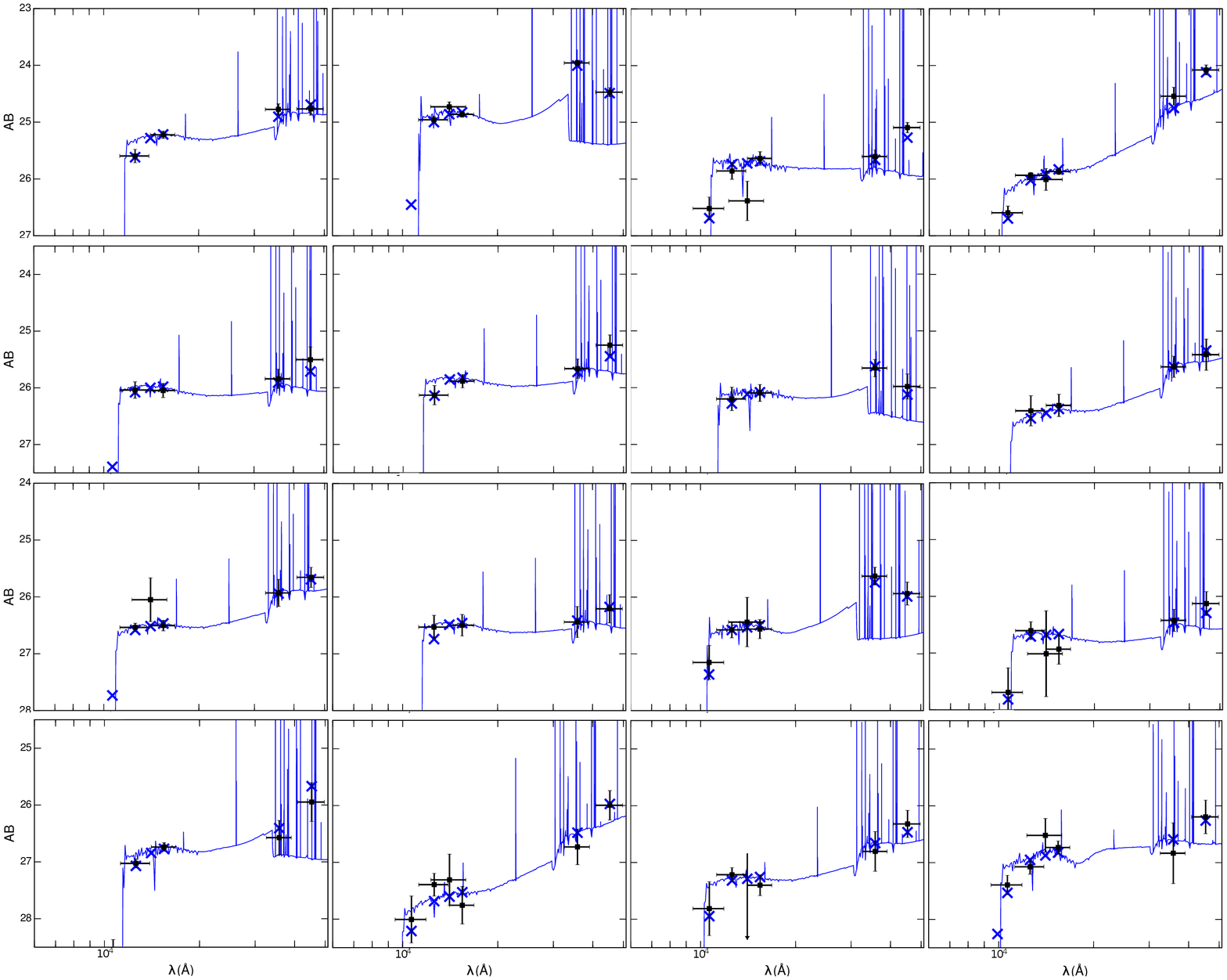}
     \caption{Examples of SEDs for our final sample. Each row shows 4 SEDs randomly selected in 4 bins of F125W magnitude defined from top to bottom as $\mathrm{F125W}<26$,  $26<\mathrm{F125W}<26.5$, $26.5<\mathrm{F125W}<27$, and $\mathrm{F125W}>27$. The  errorbars  of  the  observed wavelength  indicate the width of the filter transmission curve.  Upper  limits in flux  indicate $1\sigma$ limits. Blue crosses  show  the synthesised flux in the filters. Our grid of models is able to reproduce the large range of observed (3.6-4.5)\micron\ colors.}
\label{fig:fig3}
 \end{figure*} 

\section{Constraints on the ISM and Physical Properties}
\label{sec:ism}

\subsection{Predictions for the (3.6-4.5)\micron\ Color}

We show in the top left panel of Fig.~\ref{fig:fig2} the range of (3.6-4.5)\micron\ color which is spanned by our grid of models for $Z=0.004$. At an age of 1 Myr, (3.6-4.5)\micron\ can vary by 2 magnitudes, nebular emission (mainly \oiiidoub+\hb\ lines) boosting the flux in the IRAC 4.5\micron\ channel and producing IRAC 3.6\micron$-$IRAC 4.5\micron\ color redder by $\sim1.5$ magnitude in comparison with the color expected from a pure stellar template. However, nebular emission can also have the opposite effect for low ionization parameters, producing a bluer color than expected for pure stellar emission up to $\sim0.5$ magnitude. This effect is due to the relation between the ionization parameter and the \oiii/\oiilam\ ratio: $\log(\oiii/\oiilam)$ is increasing with higher ionization parameter \citep[depending on the metallicity,][]{kewleydopita02}, becoming larger than 1 at $Z=0.05\zsun$ for $\log\mathrm{U}\gtrsim-3.0$. 
However, the number of galaxies for which the IRAC 3.6\micron$-$IRAC 4.5\micron\ color is best fitted with a very low ionization parameter ($\log\mathrm{U}=-4.0$) is small (25\%). Furthermore, photometric redshift uncertainties can also account for $z>7.1$ blue colors with the Balmer jump starting to enter the IRAC 3.6\micron\ channel at $z>8$.

We show the EW(\oiiidoub+\hb) evolution with age in Fig.~\ref{fig:fig2} (bottom left panel) and examples of SEDs for our final sample for a range of F125W magnitude in Fig.~\ref{fig:fig3}.

\subsection{Stellar Metallicity and ISM Physical Properties}

The models required to reproduce the SEDs, mainly the IRAC colors, give insight in the ISM physical properties at $z\sim8$. Some parameters, like the C/O ratio or the hydrogen gas density, have little to no impact on the EW(\oiiidoub+\hb) which is the spectral feature with the largest effect on the IRAC colors, and therefore providing the main constraints on the ISM physical conditions. In our grid of models, the (3.6-4.5)\micron\ is mostly defined by the stellar metallicity, the ionization parameter, and the age of the stellar population. For our sample, we find a median stellar metallicity of $Z_\star=0.004_{-0.002}^{+0.004}$, a median ionization parameter $\log\mathrm{U}=-3.0\pm1.0$, and a median $\log(\mathrm{age}/\mathrm{yr})=7.2^{+0.9}_{-0.6}$. The constraints on dust extinction are mostly coming from the fit of the UV $\beta$ slope and we find a median extinction $A_V=0.4\pm0.2$. 

As noted previously, we do not assume any ISM properties of $z\sim8$ galaxies but compare their photometry with a grid of photoionization models that we consider to encompass the plausible $z\sim8$ properties. The constraints on the stellar metallicity and the ionization parameter are driven by the number of ionizing photons available to interact with the gas and the gas-phase Oxygen abundance. The ionizing photon flux is increasing with decreasing metallicity \citep[e.g.,][]{stanway+16} and with increasing ionization parameter, but the gas-phase Oxygen abundance is decreasing with decreasing metallicity, and also decreasing with increasing dust-to-metal ratio \citep[more Oxygen depletes into the ISM dust-phase;][]{gutkin+16}. Therefore the ISM parameter derivation suffers from several degeneracies: to reproduce (3.6-4.5)\micron\ colors produced by strong \oiiidoub+\hb\ emission lines ($(3.6-4.5)\micron>0$, Fig.~\ref{fig:fig2}) a large range of metallicities and ionization parameters is allowed, as long as there is the right balance between the number of ionizing photons and the Oxygen abundance. However, the parameter space allowing this balance is smaller for low metallicities due to large ionizing photon production but low Oxygen abundance. The same is true for high metallicities due to lower ionizing photon production and high Oxygen abundance. Then the median metallicity found for our sample only reflects that for $Z=0.004$ there is a larger parameter space in terms of ionization parameter, age, and dust-to-metal ratio allowing to reproduce the observed (3.6-4.5)\micron\ colors. Indeed, we found that $\sim40\%$ of our sample has a best-fit SED with $Z\leq0.002$, while the median of the marginalized likelihood for the metallicity is $Z\geq0.004$ for the entire sample.

The choice of an IMF upper mass cutoff at 300\msun\ has a negligible impact on our results since a cutoff of 100\msun\ changes the \oiiidoub\ flux by $\sim10\%$ (1-2\% for \hb\ and \oii) which is small compared to the typical uncertainties affecting EWs and line luminosities (Sec.~\ref{sec:sed}). Changing the assumed dust attenuation curve from a Calzetti to an SMC curve \citep{prevot+84,bouchet+85} has also little to no impact on the overall derived properties, except for dust attenuation.

The $z\sim8$ subsample with $S/N(3.6\mu\mathrm{m}\lor4.5\mu\mathrm{m})\geq3$ has a median stellar mass $\log(\mstar/\msun)=8.62^{+0.43}_{-0.39}$ and a median SFR $\log(\mathrm{SFR}/\msunyr)=1.26^{+0.42}_{-0.30}$.

   \begin{figure}
          \centering
     \includegraphics[width=\columnwidth,trim=0.75cm 0.5cm 2cm 1.25cm,clip=true]{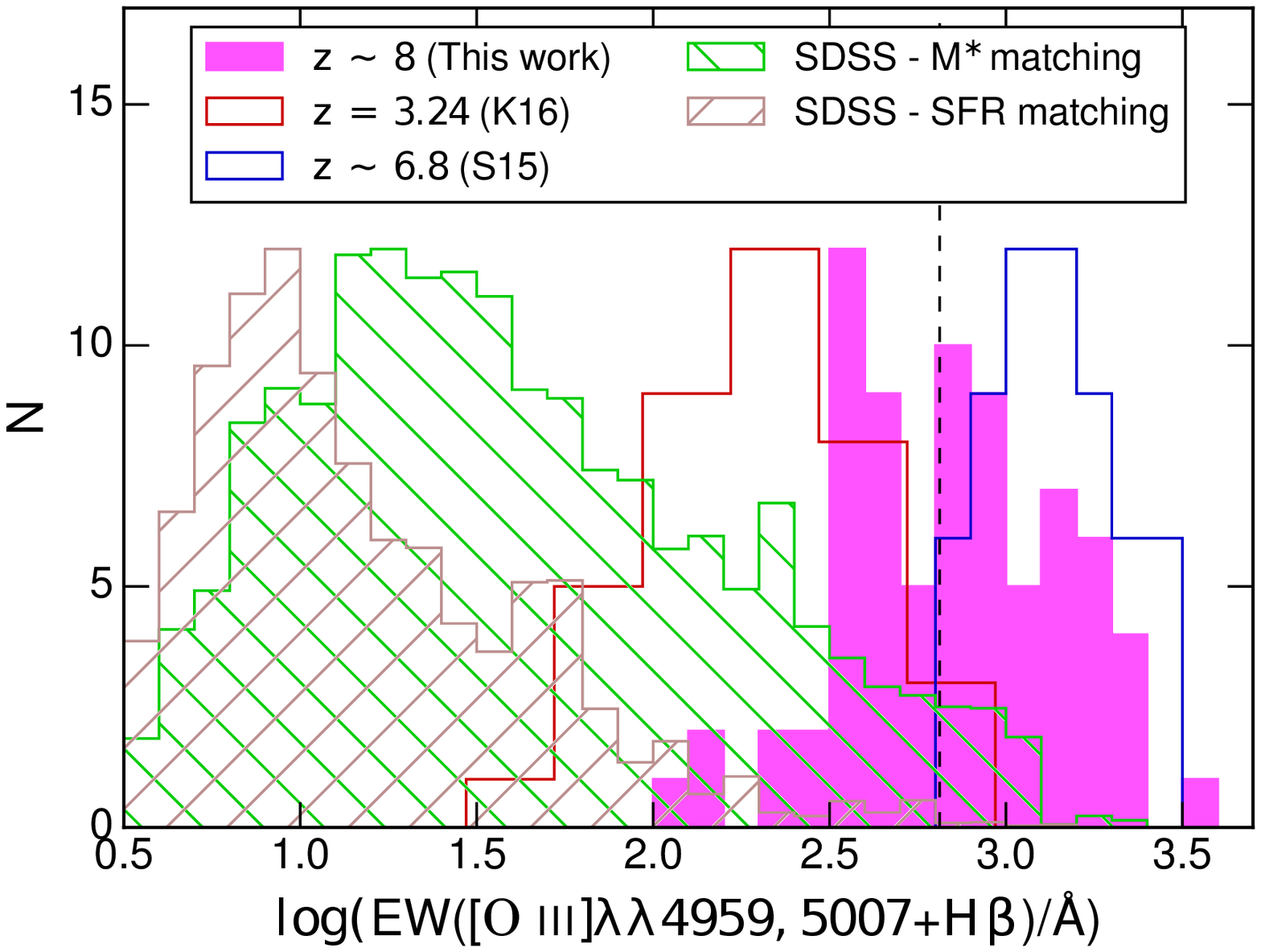}
     \caption{EW(\oiiidoub+\hb) distribution for our sample. We also show the EW distribution at $z=3.24$ \citep[][K16]{khostovan+16},  $z\sim6.8$ \citep[][S15]{smit+15}, and the distribution drawn from the SDSS sample by matching the stellar mass and SFR distributions of our $z\sim8$ sample. All distributions have been renormalized to have the same maximum as the $z\sim8$ distribution. The vertical dashed line shows the median value for our sample. Clearly, the average equivalent width of our $z\sim8$ sample is significantly higher than the $z\sim0$ or $z\sim3$ samples.  }
     \label{fig:fig4}
 \end{figure} 
 
 \subsection{Ionizing Photon Production Efficiency}
 
We are able to derive the ionizing photon production efficiency from SED fitting by computing for each template used in this work the Lyman continuum photon production rate $N_\mathrm{LyC}$ and the observed monochromatic UV luminosity $L_\nu$. The ionizing photon production  is then $\xi_\mathrm{ion}=N_\mathrm{LyC}/L_\nu$. For our final sample, we find $\log(\xi_\mathrm{ion}/\mathrm{erg}^{-1}\hspace{1mm}\mathrm{Hz})=26.29^{+0.40}_{-0.38}$ assuming a Calzetti dust attenuation curve and $\log(\xi_\mathrm{ion}/\mathrm{erg}^{-1}\hspace{1mm}\mathrm{Hz})=26.07^{+0.27}_{-0.30}$ assuming an SMC dust attenuation curve. In most lower redshift studies \citep[e.g.,][]{shivaei+18}, $\xi_\mathrm{ion}$ is inferred from a dust corrected Hydrogen line (e.g., \ha) for which the flux depends mostly on the Lyman continuum photon production rate \citep[e.g.,][]{storeyhummer95}. Given the large number of unconstrained parameters going into our analysis (e.g., attenuation curve), we consider that our result set a lower limit to the average $z\sim8$  ionizing photon production efficiency with $\log(\xi_\mathrm{ion}/\mathrm{erg}^{-1}\hspace{1mm}\mathrm{Hz})\geq25.77$.

{We note that in some studies \citep[e.g.,][]{bouwens+16a}, $\xi_\mathrm{ion}$ is an {\it intrinsic} quantity since it is derived by using a dust-corrected UV luminosity, while in our work we derive an {\it observed} $\xi_\mathrm{ion}$ value since we do not correct the observed UV luminosity for dust. Due to this dust correction, the intrinsic $\xi_\mathrm{ion}$ sets a lower limit for the observed $\xi_\mathrm{ion}$. However, thanks to the small dust attenuation that we find for our sample, the difference between intrinsic and observed $\xi_\mathrm{ion}$ should be small.

The constraints that we obtain on the ionizing photon production efficiency at $z\sim8$ are consistent with results obtained for the bluest (i.e., least dust attenuated) LBGs and Lyman-$\alpha$ Emitters at $z\sim2$ \citep{shivaei+18,sobral+18,tang+18} as well as results obtained for low-redshift compact star-forming galaxies \citep{izotov+17}, and for LBGs at $z\sim4-5$ \citep{bouwens+16a,lam+19,ceverino+19}. The observed ionizing photon production efficiency that we find is also consistent with the observed value found for $z\sim0.3$ Lyman continuum emitters \citep[$\xi_\mathrm{ion}=25.6-26$,][]{schaerer+16}.

This is the first time that the ionizing photon production efficiency is estimated for a significant sample of galaxies in the reionization era and  this value is higher than the canonical value $\log(\xi_\mathrm{ion}/\mathrm{erg}^{-1}\hspace{1mm}\mathrm{Hz})=25.2-25.3$ by a factor $\geq3$. This higher value of $\xi_\mathrm{ion}$ translates into a lower value of the Lyman continuum escape fraction required in a scenario where star-forming galaxies are driving cosmic reionization \citep{bouwens+16a,shivaei+18,chevallard+18b,matthee+17,matthee+17b,lam+19}.

 \subsection{Evolution of the EW(H$\beta$+{[O\,{\normalsize \textrm{III}}]}$\lambda\lambda4959,5007$) with Redshift}

We  show the EW(\oiiidoub+\hb) distribution for our $z\sim8$ sample in Fig.~\ref{fig:fig4} along with the EW distribution at $z=3.24$ \citep{khostovan+16} and the one at $z\sim6.8$ for the extreme emitter sample of \cite{smit+15}. Our distribution is consistent with the latter, given that the sample of \cite{smit+15} only included sources with the largest EW. We find a median 
$\mathrm{EW}(\oiiidoub+\hb)=649^{+92}_{-49}$\AA, consistent with the value of 670$^{+260}_{-170}$\AA\ from \cite{labbe+13}. 

Comparing our EW distribution with the Sloan Digital Sky Survey \citep[SDSS,][]{sdss18}, we find that only $0.23\pm0.01\%$ galaxies exhibit such strong emission lines ($\mathrm{EW}(\oiiidoub+\hb)\geq300$\AA) in the entire SDSS sample.
We also compare the $z\sim8$ distribution with two distributions drawn from SDSS.  The SDSS samples were mass- or SFR-matched by randomly picking 50 SDSS galaxies within 0.05 dex and 0.1 dex in terms of stellar mass and SFR, respectively, for each galaxy in our sample. The SDSS sample selected through SFR-matching exhibits only a small overlap with the EWs derived in our work ($1.6\pm0.2\%$), while the one matched by stellar mass leads to a non-negligible fraction of galaxies with EW as high as found in our $z\sim8$ sample ($9.8\pm0.4\%$). Nevertheless, it is clear that the emission lines of $z\sim8$ galaxies are much more extreme than local galaxies of similar mass or SFR, although specific local and low-z population such as Blue Compact Dwarf galaxies exhibit similar properties \citep[e.g.,][]{izotov+11,cardamone+09,yang+17b,rigby+15,senchyna+17}.

    \begin{figure}
          \centering
     \includegraphics[width=\columnwidth,trim=0.25cm 0cm 2cm 1.25cm,clip=true]{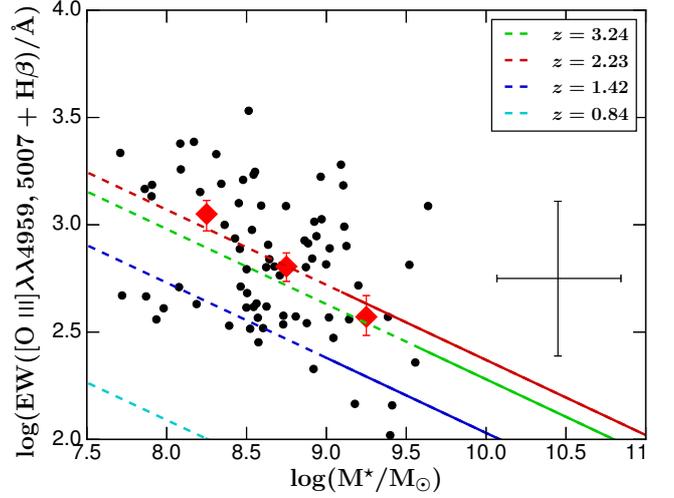}
     \caption{EW vs. \mstar\ for our $z\sim8$ sample. The large red diamonds show the median EW in bins of \mstar. We also show the power law relations $\mathrm{EW}\propto\mstar^\beta$ derived for different redshifts in \citet{khostovan+16} down to the minimum stellar mass used to derive them (solid lines) and their extrapolations to lower stellar masses (dashed lines). The median error bar for individual objects is shown on the right side of the figure. Our $z\sim8$ sample is broadly consistent with the $z\sim2-3$ relation.}
     \label{fig:fig5}
 \end{figure}

We have also attempted to match our sample in terms of sSFR but the number of SDSS galaxies to exhibit similarly high sSFR as in our sample (median $\mathrm{sSFR}=63^{+188}_{-55}$Gyr$^{-1}$) is extremely small ($<0.35\%$), such that no representative sSFR-matched sample could be constructed. However, the small sample of galaxies with such high sSFR does indeed exhibit EWs as large as derived in our work. This illustrates that finding a significant sample of local galaxies with properties (stellar mass, SFR, sSFR, and EW(\oiiidoub+\hb) similar to $z\sim8$ galaxy properties is a difficult task.

From $z=3.24$ to $z\sim8$, there is a clear evolution of the median of the EW distribution. However, the EWs have to be compared for a given stellar mass range \citep[e.g., $9.5<\log(\mstar/\msun)<10.0$,][]{khostovan+16}. The derived stellar mass for our sample is significantly lower than the $z\sim3.24$ sample, with only four galaxies (5\%) with $\log(\mstar/\msun)\geq9.5$. 
A comparison of the EW properties as a function of stellar mass is shown in Fig.~\ref{fig:fig5}.
While individual error bars are relatively large ($\sim\pm0.4$ dex for both parameters), our sample is consistent with the $z=3.24$ and $2.23$ EW--\mstar\ relations. Furthermore, using the median stellar mass of our sample with the $z=2.23$ relation from \cite{khostovan+16}, we predict a median equivalent width $\mathrm{EW}=712^{+70}_{-62}$\AA\ for our sample, a value  consistent with our derivation based on SED fitting.

 \begin{figure}
     \centering
     \includegraphics[width=\columnwidth,trim=0cm 0.5cm 1cm 0cm,clip=true]{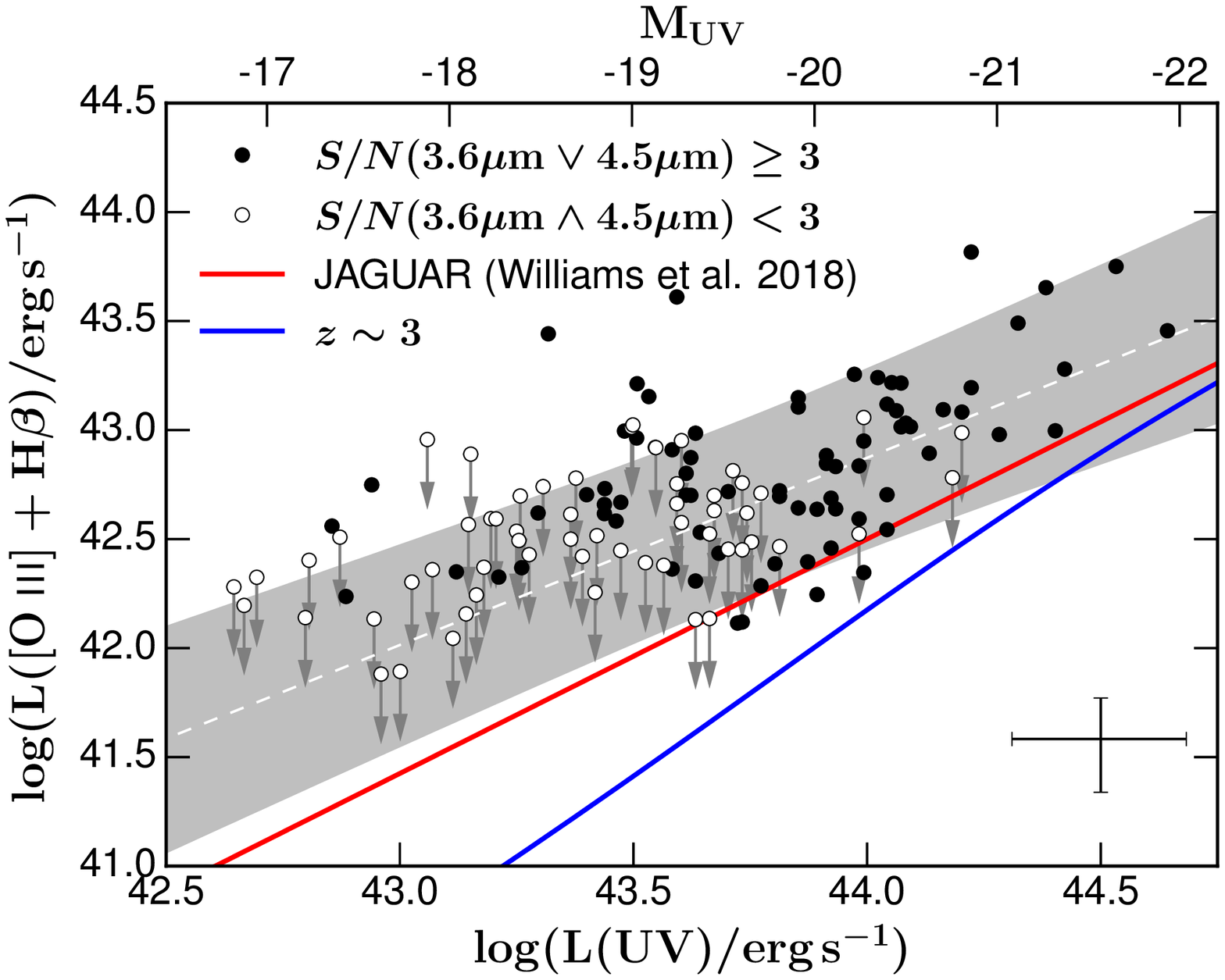}
     \caption{UV luminosity vs. \oiiidoub+\hb\ luminosity for our sample. The dashed line is the derived relation and the grey area shows the 68\% confidence interval. The typical errorbar for individual galaxies with $S/N(3.6\mu\mathrm{m}\lor4.5\mu\mathrm{m})\geq3$ is shown on the  bottom right corner. For galaxies with $S/N(3.6\mu\mathrm{m}\land4.5\mu\mathrm{m})<3$, we show the 90\% upper limits. We show in red and blue the same relation derived from the JAGUAR mock catalog \citep[][see Sec.~\ref{sec:uvoiii}]{williams+18} and the relation at $z\sim3$ (as derived through abundance matching; see Sec.~\ref{sec:hblf}), respectively. The line luminosities increase from $z\sim3$ to $z\sim8$ (at a given UV luminosity).}
     \label{fig:fig6}
 \end{figure} 

\section{The \hb+[\textrm{O}~\textsc{iii}] Luminosity Function at $z\sim8$}
\label{sec:LF}

Based on the photometric estimates of the  \oiiidoub+\hb\ emission line strengths of all the $z\sim8$ galaxies in the GREATS sample, we have the opportunity for a first derivation of the \oiiidoub+\hb\ luminosity function at these redshifts, which we describe in the next section.

\subsection{Derivation of the  Emission Line Luminosity Function}
\label{sec:uvoiii}

Our approach is based on converting the UV LF to an emission line LF using the relation between the observed UV luminosity, $L_{UV}$, and the \oiiidoub+\hb\ line luminosity, $L_{OIII+H\beta}$, at $z\sim8$. This approach is analogous to the one used in the derivation of stellar mass functions at very high redshift \citep[e.g., in][]{gonzalezetal2010,song+16} or the star-formation rate function \citep[][]{smit+12,smit+16,mashian+15}.

  \begin{figure*}
     \centering
     \includegraphics[width=\textwidth,trim=0cm 0cm 0cm 1cm,clip=true]{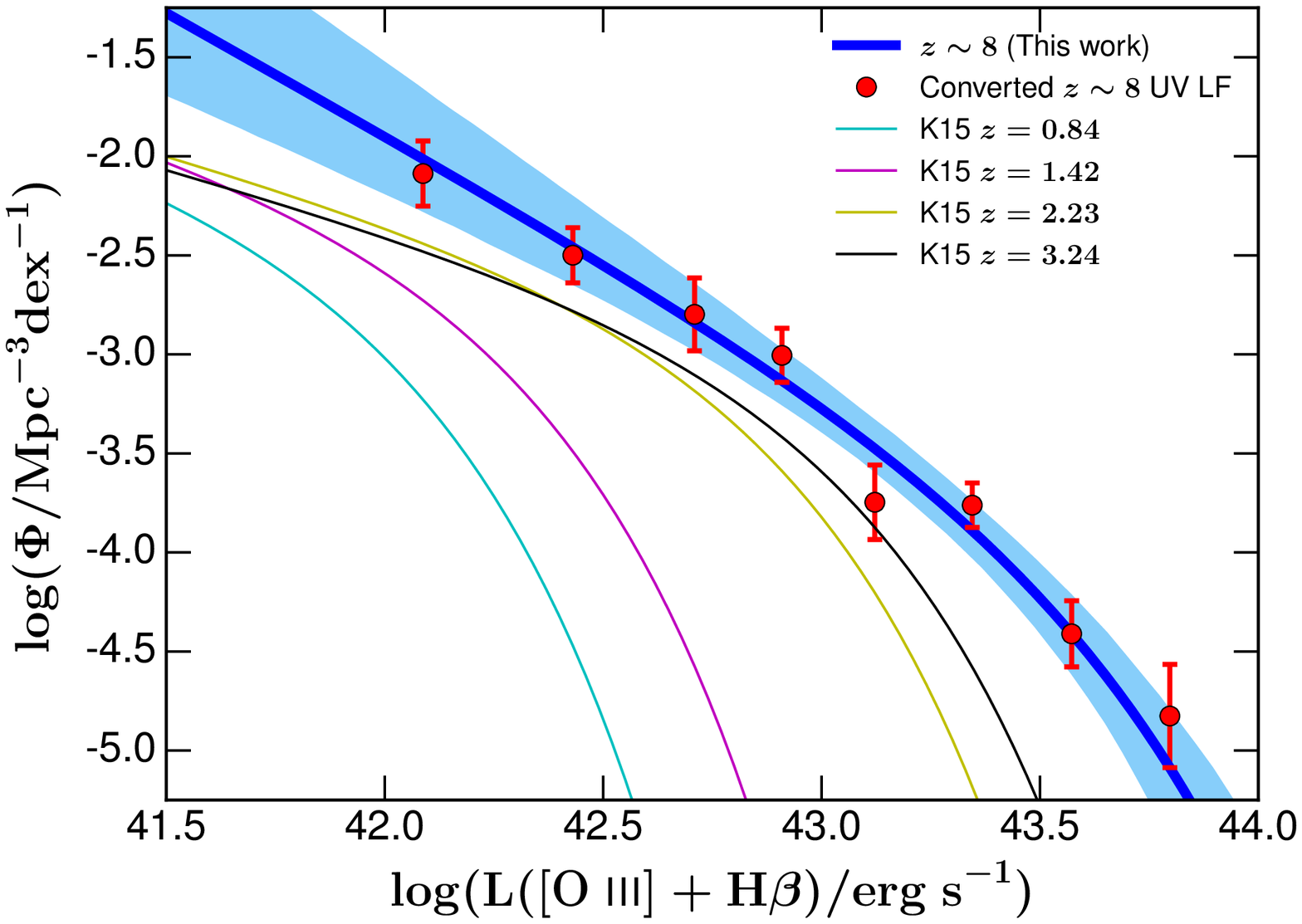}
     \caption{\oiiidoub+\hb\ luminosity function derived in this work for our sample at $z\sim8$ (blue thick line) with 68\% confidence interval (light blue area). We also show the corresponding $z\sim8$ UV luminosity function converted with the relation from Fig.~\ref{fig:fig6}. For comparison, we show the \oiiidoub+\hb\ luminosity functions measured in \citet[K15]{khostovan+15} from $z=0.84$ to $z=3.24$. Continuing the trend from lower redshift, the line LF is higher at $z\sim8$ than at $z\sim3$ (unlike the UV LF).}
     \label{fig:fig7}
 \end{figure*} 

The relation between the \oiiidoub+\hb\ luminosity and the observed UV luminosity in our sample is calibrated in Fig.~\ref{fig:fig6}. To increase the range of UV luminosities probed in our work, we add to our sample with $S/N(3.6\mu\mathrm{m}\lor4.5\mu\mathrm{m})\geq3$ galaxies with lower $S/N$. We apply to these galaxies the same procedure as the rest of the sample and so we obtain the complete probability distribution function for all the parameters, including the \oiii+\hb\ luminosity. While uncertainties remain relatively large for individual galaxies ($\sim0.18$ dex and $0.19-0.24$ dex for the UV and \oiiidoub+\hb\ luminosities, respectively), we find that the UV luminosity and the \oiii+\hb\ luminosity are well correlated (Spearman rank correlation coefficient $\rho=0.56$, standard deviation from null hypothesis $\sigma=7.1$).

We use a MCMC method to fit the relation with three parameters, a slope and an intercept, plus an intrinsic (Gaussian) dispersion around the relation, $\sigma_{int}$. This results in:
\begin{eqnarray}
\label{eq:uvoiii}
\log(\mathrm{L(\oiiidoub+\hb)}/\mathrm{erg}\hspace{1mm}\mathrm{s}^{-1}) = 0.86\pm0.12\times\nonumber\\\log(\mathrm{L_{UV}}/\mathrm{erg}\hspace{1mm}\mathrm{s}^{-1})+33.92^{+1.23}_{-1.27}
\end{eqnarray}
Together, with an intrinsic dispersion of $\sigma_{int}=0.35$\,dex around the median fit. The corresponding 68-percentile contours are also shown in Fig.~\ref{fig:fig6}.

As a comparison, we also use the publicly available mock catalog JAdes extraGalactic Ultradeep Artificial Realizations \citep[JAGUAR,][]{williams+18} to derive the relation between UV and \oiii+\hb\ luminosity of simulated $z\sim8$ galaxies. The JAGUAR mock catalog has been produced by matching luminosity and stellar mass functions as well as the relation between the stellar mass and UV luminosity, mostly at $z\leq4$. The galaxy properties are then extrapolated up to $z\sim15$. The JAGUAR catalog provides emission line fluxes and EWs for the main lines based on modeling with the \textsc{BEAGLE} code \citep{chevallard+16,chevallard+18}. We identify all galaxies from the fiducial JAGUAR mock in the redshift range $7.11<z<9.05$ and we randomly select 1000 of them to match the absolute UV magnitude distribution of our sample, and then fit the UV-\oiii+\hb\ luminosity data. The result is shown in red in Fig.~\ref{fig:fig6}. Similarly to our sample, the $z\sim8$ galaxies from the JAGUAR catalog exhibit a tight relation between UV and \oiii+\hb\ luminosity (Spearman rank correlation coefficient $\rho=0.73$, standard deviation from null hypothesis $\sigma>40$). However, the mock galaxies exhibit a significantly lower \oiii+\hb\ luminosity ($\sim0.5$ dex) at a given $L_\mathrm{UV}$ compared to the relation of our galaxies. The detailed reason for this discrepancy relative to the JAGUAR mock is unclear, but one possible reason is differences in the median physical properties. For instance, while the mock galaxies exhibit (3.6-4.5)\micron\ color similar to the ones from our sample at a given UV luminosity, the average F125W-3.6\micron\ color in JAGUAR is smaller by $\sim0.3$ magnitude compared to the observed F125W-3.6\micron\ color in our sample. This means that while (3.6-4.5)\micron\ color and EW(\oiii+\hb) are on average similar between JAGUAR and our sample, the absolute \oiii+\hb\ line luminosity scales with the 3.6\micron\ flux which is larger in our sample compared to the JAGUAR mock catalog. Furthermore, JAGUAR models a small field comparatively to our data, therefore the overlap in UV luminosity is small.

  \begin{figure*}
     \centering
     \includegraphics[width=\textwidth,trim=0cm 0cm 0cm 0.5cm,clip=true]{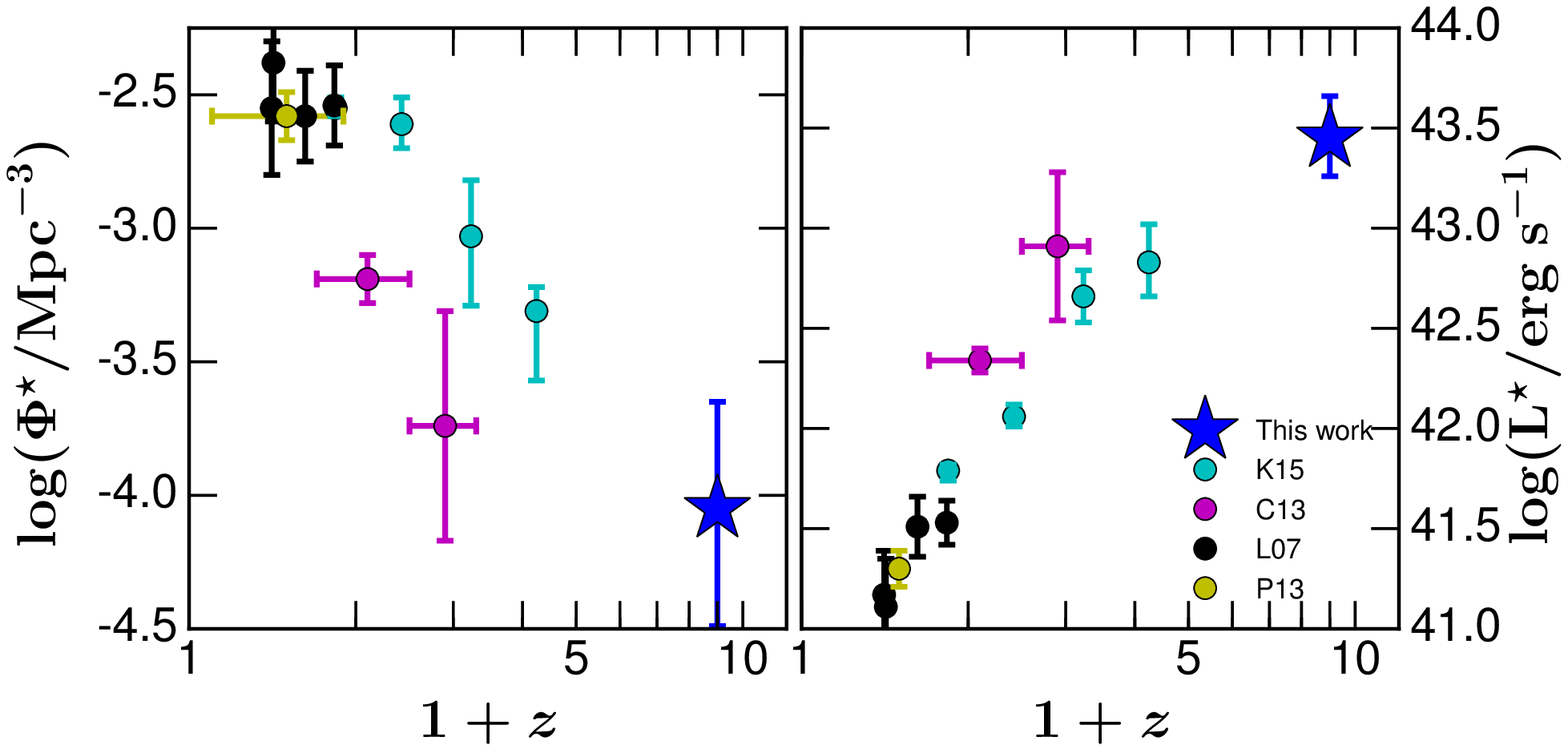}
     \caption{Evolution of $\Phi^\star$ (left panel) and $L^\star$ (right panel) with redshift with data from \citet[K15]{khostovan+15}, \citet[C13]{colbert+13}, \citet[L07]{ly+07}, and \citet[P13]{pirzkal+13}. Our values of $\Phi^\star$ and $L^\star$ agree with the extrapolated trends observed at lower redshift, indicating a relatively smooth evolution of the emission line LF.}
     \label{fig:fig8}
 \end{figure*}

Using the relation between the UV and \oiii+\hb\ luminosity of  Eq.~\ref{eq:uvoiii}, we can now derive the \oiii+\hb\ LF based on the known $z\sim8$ UV LF \citep{bouwens+15b}. In order to properly compute errorbars, we adopt a Markov Chain Monte Carlo approach \citep[see][for a review]{sharma17}. In particular, we sample 100'000 points from the UV LF and convert their corresponding $L_{UV}$ values to $L_{OIII+H\beta}$, based on the relation derived above including the appropriate dispersion $\sigma_{int}$. Finally, we fit a Schechter function to the resulting $L_{OIII+H\beta}$ values, keeping the three quantities $\Phi^\star$, $\mathrm{L}^\star$, and $\alpha$ as free parameters. We repeat this procedure 10'000 times, and vary the input Schechter function parameters of the UV LF according to their appropriate covariance matrix, which was derived from the contour plots shown in \citet{bouwens+15b}. This procedure results in 10'000 line LFs, from which we compute the mean and standard deviation for all three Schechter function parameters of the line LF.

The final result is shown in Fig.~\ref{fig:fig7}, with  Schechter function parameters of the line LF of $\log(\mathrm{L}^\star/\mathrm{erg}\hspace{1mm}\mathrm{s}^{-1})=43.45^{+0.21}_{-0.19}$, $\log(\Phi^\star/\mathrm{Mpc}^{-3})=-4.05^{+0.40}_{-0.44}$, and $\alpha=-2.22^{+0.28}_{-0.32}$. The points with errorbars in Fig.~\ref{fig:fig7} correspond to the observed UV LF measurements from \citet{bouwens+15b}, which were converted to the \oiiidoub+\hb\ LF using the same approach as described above. They clearly agree very well with the mean Schechter function derivation.

\subsection{The Evolution of the $[OIII]+H\beta$ Line Luminosity Function to $z\sim8$}
\label{sec:hblf}

The \oiiidoub+\hb\ LF has previously been derived up to $z\sim3$ by several authors \citep{hippelein+03,ly+07,colbert+13,pirzkal+13,khostovan+15}. By adding our estimate at $z\sim8$, we can thus study its evolution across more than 13 Gyr. 
Fig.~\ref{fig:fig7} shows such a comparison of the $L_{OIII+H\beta}$ LFs derived at different redshift, from $z\sim0.5$ to $z\sim8$.
Interestingly, the line LF is found to be higher at all luminosities at $z\sim8$ compared to $z\sim3$. This is in stark contrast to the evolution of the UV LF, which peaks at $z\sim2-3$, but then steadily declines to higher (or lower) redshift. We show in Fig.~\ref{fig:fig8} the evolution with redshift of L$^\star$ and $\Phi^\star$ and by extrapolating the observed trends at lower redshift to $z\sim8$, especially the \cite{khostovan+15} results, the $z\sim8$ values are in remarkable agreement with expectations.

  \begin{figure}
       \centering
     \includegraphics[width=\columnwidth,trim=0.5cm 0cm 1cm 0cm,clip=true]{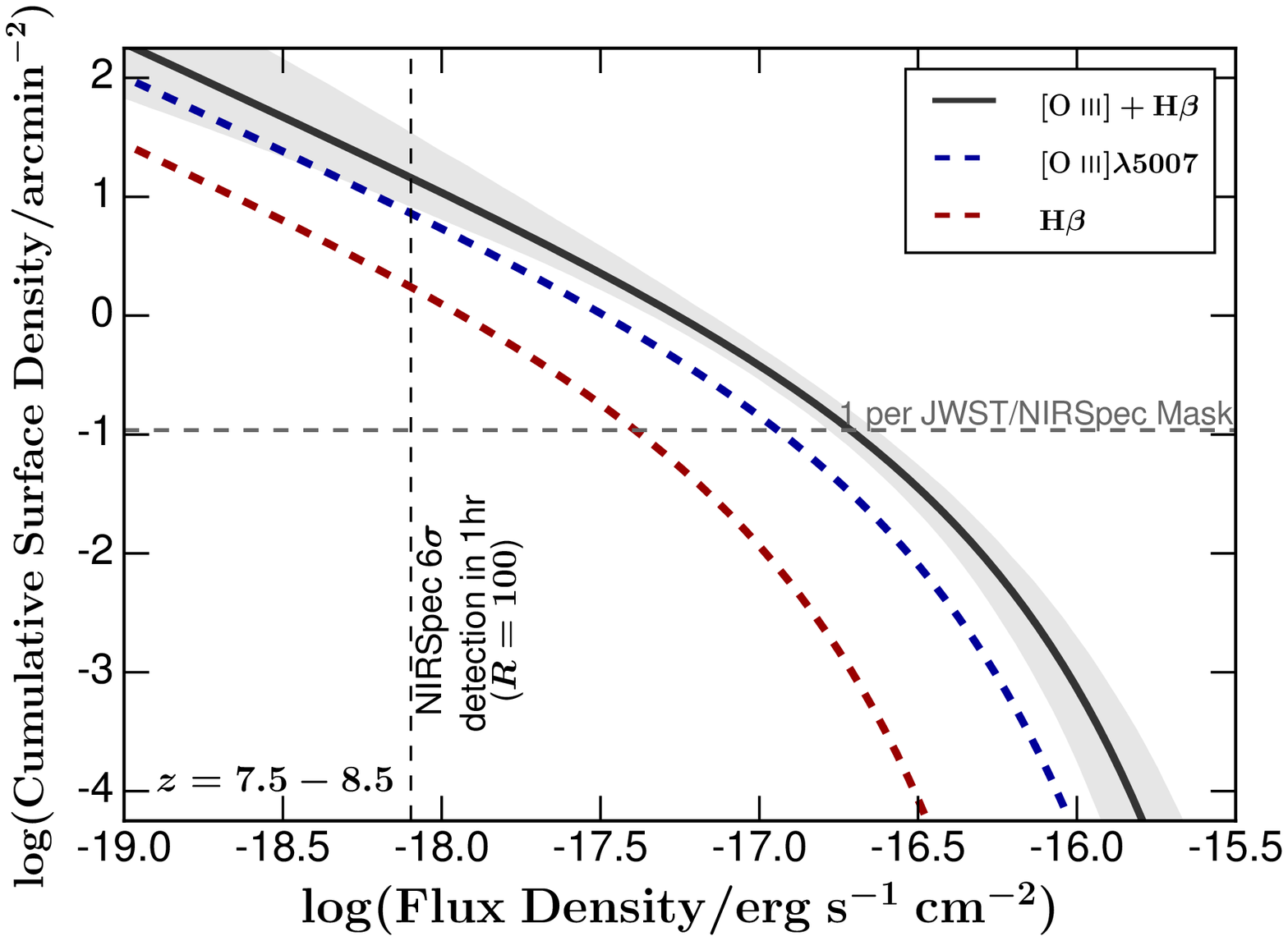}
     \caption{Cumulative surface density as a function of emission line flux densities at $z\sim8$ derived from the \oiiidoub+\hb\ luminosity function presented in this work. We also show the relations for \oiiiv\ and \hb, assuming the median contribution of each of these lines to the total \oiiidoub+\hb\ luminosity obtained from the SED fitting. We use this figure to make \jwst\ number counts prediction. Already at the 1hr depth of NIRSpec, we expect to be able to more than fill an entire mask with $z\sim8$ galaxies, but deeper pre-imaging is required to identify these targets in most fields  (see Text).}
     \label{fig:fig9}
 \end{figure}

The evolution of the \oiii+\hb\ LF from $z\sim3$ to $z\sim8$ can be explained by an evolution of the relation between L(UV) and the stellar mass with redshift. It is known that high-redshift galaxies have their stellar mass related to their UV luminosity \citep{stark+09,gonzalez+11,duncan+14,grazian+15,song+16,stefanon+17} and while at $0<z\leq4$ the slope of the $M_\mathrm{UV}-\mstar$ relation is not evolving, the intercept evolves in such way that at a given stellar mass the UV luminosity decreases with increasing redshift \citep{williams+18}. The evolution of this relation is more uncertain at $z>4$ because of the uncertainties on the stellar mass estimation due to the nebular emission contamination \citep[e.g.,][]{debarros+14}, but assuming a similar evolution at $z>4$, combining the evolution of the UV luminosity at a given stellar mass and the relation observed between nebular emission line EW and stellar mass \citep[Fig.~\ref{fig:fig5};][]{fumagalli+12,sobral+14,khostovan+16}, we expect that with increasing redshift, at a given UV luminosity, the stellar mass decreases and the EW(\oiii+\hb) increases accordingly. This means that at a given UV luminosity the \oiii+\hb\ luminosity is increasing with increasing redshift. This is the expected trend but uncertainties on the stellar mass estimation (Fig.~\ref{fig:fig5}) precludes any further quantification of the UV luminosity vs. stellar mass relation evolution from $z\sim3$ to $z\sim8$. 

To test this explanation, we derive the relation between L(\oiii+\hb) and L(UV) at $z\sim3$ through abundance matching. In particular, we use the $z\sim3$ UV LF from \cite{reddysteidel09} to derive the cumulative number density of galaxies at a given UV luminosity and match it to the corresponding cumulative number density at a given \oiii+\hb\ luminosity based on the $z=3.2$ \oiii+\hb\ LF from \cite{khostovan+15}. We show the resulting relation in Fig.~\ref{fig:fig6} with a blue line. There is a clear evolution of the L(UV) vs. L(\oiii+\hb) relation from $z\sim3$ to $z\sim8$, galaxies being brighter in \oiii+\hb\ at any given UV luminosity explored in this work and an increasing difference between the $z\sim3$ and $z\sim8$ relation with decreasing UV luminosity. This finding supports our explanation for the \oiii+\hb\ LF evolution.

\subsection{Predictions of JWST Number Counts}

One of the most awaited capabilities of the upcoming $JWST$ is its unprecedented sensitivity for spectroscopy at $>2\micron$. In particular, $JWST$ will for the first time provide spectroscopic access to the rest-frame optical emission lines of very high redshift galaxies, including the \oiiidoub\ and \hb\ lines at $z\sim8$ that we constrained through photometry here. In order to decide on the area and depth for the most efficient spectroscopic surveys with $JWST$, an estimate of the \oiiidoub+\hb\ LF as derived above is of critical importance.

Of particular interest for such $JWST$ predictions is the NIRSpec instrument \citep{nirspec}, which will be the workhorse NIR spectrograph.
With four quadrants, each of which covers 2.3 arcmin$^2$, NIRSpec spans an effective area of $\sim9.2$ arcmin$^2$. Its sensitivity is exquisite. In only 1hr, NIRSpec will reach $6\sigma$ detections for emission lines at $\sim$4.5$\micron$ and fluxes of $\sim$6$\times10^{-19}$\,erg\,s$^{-1}$cm$^{-2}$ at $R=1000$, or $\sim$8$\times10^{-19}$\,erg\,s$^{-1}$cm$^{-2}$ at $R=100$. These numbers were derived with the latest \jwst/ETC, when integrating over the full extent of the lines (which were assumed to have an intrinsic width of 150 km s$^{-1}$).

In Fig.~\ref{fig:fig9}, we plot the cumulative surface density of galaxies at $z=7.5-8.5$ as a function of emission line luminosities based on the line LF which we derived in the previous section. In particular, we show the total combined \oiiidoub+\hb\ luminosities (as would be seen, e.g., in $R=100$ low-resolution spectroscopy with NIRSpec), and we also split the luminosities into the three different lines, $H\beta$, [OIII]4959, and [OIII]5007. For the latter step, we employ the median line ratios as found in the SED fitting in Section 4, with \hb/(\hb+\oiiidoub)=0.21, \oiiiiv/(\hb+\oiiidoub)=0.20, and \oiiiv/(\hb+\oiiidoub)=0.59.

As can be seen, at the 1hr sensitivity limits of NIRSpec with $R=100$, we can expect a cumulative surface density of $z=7.5-8.5$ galaxies of 10$^{1.2\pm0.3}$ arcmin$^{-2}$ which have blended [OIII]$+H\beta$ lines that are bright enough to be significantly detected. This means that a single NIRSpec mask (with effective area 9.2 arcmin$^2$) would, in principle, be able to target on average $\sim150$ galaxies (where the $1\sigma$ uncertainties range from 80 to 300 galaxies). The fixed grid of the NIRSpec slitlet masks will reduce this number somewhat. 
Unfortunately, however, there is an additional limitation of early spectroscopic surveys. The depth reached in terms of \oiii+\hb\ luminosity in 1hr for NIRSpec corresponds to an observed UV magnitude of $m_\mathrm{UV}=29.9$ ($\sim29-31$ when accounting for the uncertainties in the UV-\oiii+\hb\ luminosity relation, Eq.~\ref{eq:uvoiii}). This implies that the average surface density of current $z\sim8$ galaxy samples in the prime extragalactic legacy fields such as CANDELS is significantly lower than the number above \citep{bouwens+15b}. Therefore, early \jwst\ spectroscopic surveys, which are based on the selection of targets from current $HST$ datasets, cannot be maximally efficient for a targeted $z\sim8$ galaxy survey. The best strategy will be to perform deep pre-imaging with \jwst\ to identify targets at $z\sim8$.

Nevertheless, our calculation shows that if significantly deep imaging data are available to select targets from, a single NIRSpec mask with $R=100$ could be filled with just $z\sim8$ emission line sources for which a 1hr observation can measure a secure redshift. This will result in revolutionary insights of the large scale structure in the heart of the reionization epoch.

Of course, in order to study the physics, more than a simple redshift measurement is required. In particular, the [OIII]$+H\beta$ lines need to be split with observations at $R=1000$ or higher.
For such surveys, the corresponding number of expected galaxies in 1hr observations are only 24 galaxies for $H\beta$ lines, and 97 galaxies with 6$\sigma$ [OIII]5007 line detections, per NIRSpec mask.

As a final remark, we compare our predictions with the ones from the JAGUAR mock catalog. Given the lower \oiii+\hb\ luminosities at a given L(UV) compared to our observed sample, it is clear that the mocks will significantly underpredict the number of observed sources at a given line luminosity. When computing the line LF from the JAGUAR catalogs, we find a lower normalization by a factor $\sim8\times$ compared to our observed LF. Hence, the JAGUAR mock catalogs should underpredict the number of rest-frame optical emission lines that can be detected at $z\sim8$ with $JWST$/NIRSpec in the future by the same factor.

\section{Conclusions}
\label{sec:conc}

We have presented a detailed analysis of a $z\sim8$ galaxy sample with some of the deepest available \spitzer\ observations, from the GREATS survey. The sample has been culled through photometric redshifts to ensure that the selected galaxies are at $z\geq7.11$, where the IRAC 3.6\micron$-$IRAC 4.5\micron\ colors put strong constraints on the \oiii+\hb\ equivalent width. We built and used a photoionization grid with a large parameter space covering a variety of stellar metallicity and ISM conditions, using the BPASS models as stellar emission inputs. We feed the resulting SEDs that include stellar and nebular emission (continuum and lines) to our SED fitting code to derive $z\sim8$ galaxy properties. Accounting for the photometric and model uncertainties, we have specifically derived the \oiii+\hb\ luminosities, allowing us to derive the corresponding \oiii+\hb\ luminosity function and make predictions for \jwst\ observations. 

In summary, we find the following.

\begin{enumerate}
\item Our subsample with $S/N(3.6\mu\mathrm{m}\lor4.5\mu\mathrm{m})\geq3$ has the following average properties: $\log(\mstar/\msun)=8.62^{+0.43}_{-0.39}$, $\log(\mathrm{age}/\mathrm{yr})=7.2^{+0.9}_{-0.6}$, $\log(\mathrm{SFR}/\msunyr)=1.26^{+0.42}_{-0.30}$, $A_V=0.4\pm0.2$, and $\mathrm{sSFR}=63^{+188}_{-55}\mathrm{Gyr}^{-1}$.
\item To reproduce the observed IRAC color of this subsample, which is strongly affected by \oiiidoub+\hb\ emission, the two main parameters driving the EW are the stellar metallicity and the ionization parameter, and they have the following values $Z_\star=0.004_{-0.002}^{+0.004}$ and $\log\mathrm{U}=-3.0\pm1.0$.
\item We are able to put constraints on the median ionizing photon production efficiency with $\log(\xi_\mathrm{ion}/\mathrm{erg}^{-1}\hspace{1mm}\mathrm{Hz})\geq25.77$. This latter value is $>3$ times higher than the canonical value, implying that these galaxies have a higher ionizing output than typically assumed and can thus more easily reionize the universe.
\item According to our SED fitting which matches the observed IRAC colors, we find a median rest-frame equivalent width  $\mathrm{EW}(\oiiidoub+\hb)=649^{+92}_{-49}$\AA\ (Fig.~\ref{fig:fig5}).
\item We find a relatively tight relation between \oiii+\hb\ and UV luminosity (Fig.~\ref{fig:fig6}), allowing us to derive for the first time the \oiii+\hb\ LF at $z\sim8$ based on the $z\sim8$ UV LF. We find that, in contrast with the evolution of the UV LF from $z\sim3$ to $z\sim8$, the $z\sim8$ \oiii+\hb\ LF is higher at all luminosities than at $z\sim3$. This is due to the increasing \oiii+\hb\ luminosity at a given UV luminosity with increasing redshift. 
\item Finally, we use the derived \oiii+\hb\ LF to predict \jwst\ number counts. A single NIRSpec pointing would contain $\sim150$ galaxies at $z=7.5-8.5$, for which the \oiii+\hb\ emission could be detected in only 1hr.
However, the current average surface density of $z\sim8$ galaxies in extragalactic legacy fields is significantly lower that this number. Therefore, to maximize the efficiency of \jwst\, deep pre-imaging to $m_\mathrm{UV}\sim30$ mag will be required.
\end{enumerate}

In this work, we have used the deepest \spitzer\ data available on large areas. We have accounted for observational uncertainties on the photometry and we have used a grid of photoionization models with a large parameter space. While we have attempted to minimize the number of assumptions going into our analysis, many modeling uncertainties are still present: the ingredients in the stellar population synthesis models (stellar atmospheres, binaries, rotation), the IMF, the possible presence of multiple stellar populations, the dust attenuation curve, the ratio between nebular and stellar attenuation, interstellar abundances, and depletion factor of metals on to dust grains. Only the unprecedented abilities of \jwst\ will allow to alleviate some of these uncertainties.

\section*{Acknowledgements}
We thank the anonymous referee who helped improve this manuscript. The work of SDB has been partially supported by a Flexibility Grant from the Swiss National Science Foundation and by a MERAC Funding and Travel Award from the Swiss Society for Astrophysics and Astronomy. V. G. was supported by CONICYT/FONDECYT initiation grant number 11160832.

This work made use of v2.1 of the Binary Population and Spectra Synthesis (BPASS) models as last described in \cite{eldridge+17}.
Calculations were performed with version 17.00 of \cloudy\, last described by \cite{ferland+17}.
This work also made use of Astropy, a community-developed core Python package for Astronomy \citep{astropy}, as well as the pymc3 library \citep{pymc3}.

This paper made use of public catalogs derived from data taken by the Sloan Digital Sky Survey IV. Funding for the Sloan Digital Sky Survey IV has been provided by the Alfred P. Sloan Foundation, the U.S. Department of Energy Office of Science, and the Participating Institutions. SDSS-IV acknowledges
support and resources from the Center for High-Performance Computing at
the University of Utah. The SDSS web site is www.sdss.org.





\bibliographystyle{mnras}
\bibliography{ref}

\begin{thebibliography}{}
\makeatletter
\relax
\def\mn@urlcharsother{\let\do\@makeother \do\$\do\&\do\#\do\^\do\_\do\%\do\~}
\def\mn@doi{\begingroup\mn@urlcharsother \@ifnextchar [ {\mn@doi@}
  {\mn@doi@[]}}
\def\mn@doi@[#1]#2{\def\@tempa{#1}\ifx\@tempa\@empty \href
  {http://dx.doi.org/#2} {doi:#2}\else \href {http://dx.doi.org/#2} {#1}\fi
  \endgroup}
\def\mn@eprint#1#2{\mn@eprint@#1:#2::\@nil}
\def\mn@eprint@arXiv#1{\href {http://arxiv.org/abs/#1} {{\tt arXiv:#1}}}
\def\mn@eprint@dblp#1{\href {http://dblp.uni-trier.de/rec/bibtex/#1.xml}
  {dblp:#1}}
\def\mn@eprint@#1:#2:#3:#4\@nil{\def\@tempa {#1}\def\@tempb {#2}\def\@tempc
  {#3}\ifx \@tempc \@empty \let \@tempc \@tempb \let \@tempb \@tempa \fi \ifx
  \@tempb \@empty \def\@tempb {arXiv}\fi \@ifundefined
  {mn@eprint@\@tempb}{\@tempb:\@tempc}{\expandafter \expandafter \csname
  mn@eprint@\@tempb\endcsname \expandafter{\@tempc}}}

\bibitem[\protect\citeauthoryear{{Abolfathi} et~al.,}{{Abolfathi}
  et~al.}{2018}]{sdss18}
{Abolfathi} B.,  et~al., 2018, \mn@doi [\apjs] {10.3847/1538-4365/aa9e8a},
  \href {http://adsabs.harvard.edu/abs/2018ApJS..235...42A} {235, 42}

\bibitem[\protect\citeauthoryear{{Amor{\'{\i}}n} et~al.,}{{Amor{\'{\i}}n}
  et~al.}{2017}]{amorin+17}
{Amor{\'{\i}}n} R.,  et~al., 2017, \mn@doi [Nature Astronomy]
  {10.1038/s41550-017-0052}, \href
  {http://adsabs.harvard.edu/abs/2017NatAs...1E..52A} {1, 0052}

\bibitem[\protect\citeauthoryear{{Bagnasco} et~al.,}{{Bagnasco}
  et~al.}{2007}]{nirspec}
{Bagnasco} G.,  et~al., 2007, in Cryogenic Optical Systems and Instruments XII.
  p. 66920M, \mn@doi{10.1117/12.735602}

\bibitem[\protect\citeauthoryear{{Baldwin}, {Phillips}  \&
  {Terlevich}}{{Baldwin} et~al.}{1981}]{baldwin+81}
{Baldwin} J.~A.,  {Phillips} M.~M.,   {Terlevich} R.,  1981, \mn@doi [\pasp]
  {10.1086/130766}, \href {http://adsabs.harvard.edu/abs/1981PASP...93....5B}
  {93, 5}

\bibitem[\protect\citeauthoryear{{Berg}, {Erb}, {Auger}, {Pettini}  \&
  {Brammer}}{{Berg} et~al.}{2018}]{berg+18}
{Berg} D.~A.,  {Erb} D.~K.,  {Auger} M.~W.,  {Pettini} M.,   {Brammer} G.~B.,
  2018, \mn@doi [\apj] {10.3847/1538-4357/aab7fa}, \href
  {http://adsabs.harvard.edu/abs/2018ApJ...859..164B} {859, 164}

\bibitem[\protect\citeauthoryear{{Bolzonella}, {Miralles}  \&
  {Pell{\'o}}}{{Bolzonella} et~al.}{2000}]{bolzonella+00}
{Bolzonella} M.,  {Miralles} J.,   {Pell{\'o}} R.,  2000, \aap, \href
  {http://adsabs.harvard.edu/abs/2000A%26A...363..476B} {363, 476}

\bibitem[\protect\citeauthoryear{{Bouchet}, {Lequeux}, {Maurice}, {Prevot}  \&
  {Prevot-Burnichon}}{{Bouchet} et~al.}{1985}]{bouchet+85}
{Bouchet} P.,  {Lequeux} J.,  {Maurice} E.,  {Prevot} L.,   {Prevot-Burnichon}
  M.~L.,  1985, \aap, \href
  {http://adsabs.harvard.edu/abs/1985A%26A...149..330B} {149, 330}

\bibitem[\protect\citeauthoryear{{Bouwens} et~al.,}{{Bouwens}
  et~al.}{2014}]{bouwens+14}
{Bouwens} R.~J.,  et~al., 2014, \mn@doi [\apj] {10.1088/0004-637X/793/2/115},
  \href {http://adsabs.harvard.edu/abs/2014ApJ...793..115B} {793, 115}

\bibitem[\protect\citeauthoryear{{Bouwens} et~al.,}{{Bouwens}
  et~al.}{2015}]{bouwens+15b}
{Bouwens} R.~J.,  et~al., 2015, \mn@doi [\apj] {10.1088/0004-637X/803/1/34},
  \href {http://adsabs.harvard.edu/abs/2015ApJ...803...34B} {803, 34}

\bibitem[\protect\citeauthoryear{{Bouwens}, {Smit}, {Labb{\'e}}, {Franx},
  {Caruana}, {Oesch}, {Stefanon}  \& {Rasappu}}{{Bouwens}
  et~al.}{2016}]{bouwens+16a}
{Bouwens} R.~J.,  {Smit} R.,  {Labb{\'e}} I.,  {Franx} M.,  {Caruana} J.,
  {Oesch} P.,  {Stefanon} M.,   {Rasappu} N.,  2016, \mn@doi [\apj]
  {10.3847/0004-637X/831/2/176}, \href
  {http://adsabs.harvard.edu/abs/2016ApJ...831..176B} {831, 176}

\bibitem[\protect\citeauthoryear{{Brammer}, {van Dokkum}  \& {Coppi}}{{Brammer}
  et~al.}{2008}]{brammer+08}
{Brammer} G.~B.,  {van Dokkum} P.~G.,   {Coppi} P.,  2008, \mn@doi [\apj]
  {10.1086/591786}, \href {http://adsabs.harvard.edu/abs/2008ApJ...686.1503B}
  {686, 1503}

\bibitem[\protect\citeauthoryear{{Bruzual} \& {Charlot}}{{Bruzual} \&
  {Charlot}}{2003}]{BC03}
{Bruzual} G.,  {Charlot} S.,  2003, \mn@doi [\mnras]
  {10.1046/j.1365-8711.2003.06897.x}, \href
  {http://adsabs.harvard.edu/abs/2003MNRAS.344.1000B} {344, 1000}

\bibitem[\protect\citeauthoryear{{Calzetti}, {Armus}, {Bohlin}, {Kinney},
  {Koornneef}  \& {Storchi-Bergmann}}{{Calzetti} et~al.}{2000}]{calzetti+00}
{Calzetti} D.,  {Armus} L.,  {Bohlin} R.~C.,  {Kinney} A.~L.,  {Koornneef} J.,
   {Storchi-Bergmann} T.,  2000, \mn@doi [\apj] {10.1086/308692}, \href
  {http://adsabs.harvard.edu/abs/2000ApJ...533..682C} {533, 682}

\bibitem[\protect\citeauthoryear{{Cardamone} et~al.,}{{Cardamone}
  et~al.}{2009}]{cardamone+09}
{Cardamone} C.,  et~al., 2009, \mn@doi [\mnras]
  {10.1111/j.1365-2966.2009.15383.x}, \href
  {http://adsabs.harvard.edu/abs/2009MNRAS.399.1191C} {399, 1191}

\bibitem[\protect\citeauthoryear{{Cardelli}, {Clayton}  \& {Mathis}}{{Cardelli}
  et~al.}{1989}]{cardelli+89}
{Cardelli} J.~A.,  {Clayton} G.~C.,   {Mathis} J.~S.,  1989, \mn@doi [\apj]
  {10.1086/167900}, \href {http://adsabs.harvard.edu/abs/1989ApJ...345..245C}
  {345, 245}

\bibitem[\protect\citeauthoryear{{Ceverino}, {Klessen}  \& {Glover}}{{Ceverino}
  et~al.}{2019}]{ceverino+19}
{Ceverino} D.,  {Klessen} R.~S.,   {Glover} S.~C.~O.,  2019, \mn@doi [\mnras]
  {10.1093/mnras/stz079}, \href
  {http://adsabs.harvard.edu/abs/2019MNRAS.484.1366C} {484, 1366}

\bibitem[\protect\citeauthoryear{{Chary}, {Stern}  \& {Eisenhardt}}{{Chary}
  et~al.}{2005}]{chary+05}
{Chary} R.-R.,  {Stern} D.,   {Eisenhardt} P.,  2005, \mn@doi [\apjl]
  {10.1086/499205}, \href {http://adsabs.harvard.edu/abs/2005ApJ...635L...5C}
  {635, L5}

\bibitem[\protect\citeauthoryear{{Chevallard} \& {Charlot}}{{Chevallard} \&
  {Charlot}}{2016}]{chevallard+16}
{Chevallard} J.,  {Charlot} S.,  2016, \mn@doi [\mnras]
  {10.1093/mnras/stw1756}, \href
  {http://adsabs.harvard.edu/abs/2016MNRAS.462.1415C} {462, 1415}

\bibitem[\protect\citeauthoryear{{Chevallard} et~al.,}{{Chevallard}
  et~al.}{2018a}]{chevallard+18}
{Chevallard} J.,  et~al., 2018a, \mn@doi [\mnras] {10.1093/mnras/sty2426},
  \href {http://adsabs.harvard.edu/abs/2018MNRAS.tmp.2308C} {}

\bibitem[\protect\citeauthoryear{{Chevallard} et~al.,}{{Chevallard}
  et~al.}{2018b}]{chevallard+18b}
{Chevallard} J.,  et~al., 2018b, \mn@doi [\mnras] {10.1093/mnras/sty1461},
  \href {http://adsabs.harvard.edu/abs/2018MNRAS.479.3264C} {479, 3264}

\bibitem[\protect\citeauthoryear{{Colbert} et~al.,}{{Colbert}
  et~al.}{2013}]{colbert+13}
{Colbert} J.~W.,  et~al., 2013, \mn@doi [\apj] {10.1088/0004-637X/779/1/34},
  \href {http://adsabs.harvard.edu/abs/2013ApJ...779...34C} {779, 34}

\bibitem[\protect\citeauthoryear{{De Barros}, {Schaerer}  \& {Stark}}{{De
  Barros} et~al.}{2014}]{debarros+14}
{De Barros} S.,  {Schaerer} D.,   {Stark} D.~P.,  2014, \mn@doi [\aap]
  {10.1051/0004-6361/201220026}, \href
  {http://adsabs.harvard.edu/abs/2014A%26A...563A..81D} {563, A81}

\bibitem[\protect\citeauthoryear{{De Barros}, {Reddy}  \& {Shivaei}}{{De
  Barros} et~al.}{2016}]{debarros+16b}
{De Barros} S.,  {Reddy} N.,   {Shivaei} I.,  2016, \mn@doi [\apj]
  {10.3847/0004-637X/820/2/96}, \href
  {http://adsabs.harvard.edu/abs/2016ApJ...820...96D} {820, 96}

\bibitem[\protect\citeauthoryear{{Dom{\'{\i}}nguez} et~al.,}{{Dom{\'{\i}}nguez}
  et~al.}{2013}]{dominguez+13}
{Dom{\'{\i}}nguez} A.,  et~al., 2013, \mn@doi [\apj]
  {10.1088/0004-637X/763/2/145}, \href
  {http://adsabs.harvard.edu/abs/2013ApJ...763..145D} {763, 145}

\bibitem[\protect\citeauthoryear{{Duncan} et~al.,}{{Duncan}
  et~al.}{2014}]{duncan+14}
{Duncan} K.,  et~al., 2014, \mn@doi [\mnras] {10.1093/mnras/stu1622}, \href
  {http://adsabs.harvard.edu/abs/2014MNRAS.444.2960D} {444, 2960}

\bibitem[\protect\citeauthoryear{{Eldridge}, {Izzard}  \& {Tout}}{{Eldridge}
  et~al.}{2008}]{elrdidge+08}
{Eldridge} J.~J.,  {Izzard} R.~G.,   {Tout} C.~A.,  2008, \mn@doi [\mnras]
  {10.1111/j.1365-2966.2007.12738.x}, \href
  {http://adsabs.harvard.edu/abs/2008MNRAS.384.1109E} {384, 1109}

\bibitem[\protect\citeauthoryear{{Eldridge}, {Stanway}, {Xiao}, {McClelland},
  {Taylor}, {Ng}, {Greis}  \& {Bray}}{{Eldridge} et~al.}{2017}]{eldridge+17}
{Eldridge} J.~J.,  {Stanway} E.~R.,  {Xiao} L.,  {McClelland} L.~A.~S.,
  {Taylor} G.,  {Ng} M.,  {Greis} S.~M.~L.,   {Bray} J.~C.,  2017, \mn@doi
  [\pasa] {10.1017/pasa.2017.51}, \href
  {http://adsabs.harvard.edu/abs/2017PASA...34...58E} {34, e058}

\bibitem[\protect\citeauthoryear{{Ellis} et~al.,}{{Ellis}
  et~al.}{2013}]{ellis+13}
{Ellis} R.~S.,  et~al., 2013, \mn@doi [\apjl] {10.1088/2041-8205/763/1/L7},
  \href {http://adsabs.harvard.edu/abs/2013ApJ...763L...7E} {763, L7}

\bibitem[\protect\citeauthoryear{{Faisst} et~al.,}{{Faisst}
  et~al.}{2016}]{faisst+16}
{Faisst} A.~L.,  et~al., 2016, \mn@doi [\apj] {10.3847/0004-637X/821/2/122},
  \href {http://adsabs.harvard.edu/abs/2016ApJ...821..122F} {821, 122}

\bibitem[\protect\citeauthoryear{{Ferland} et~al.,}{{Ferland}
  et~al.}{2017}]{ferland+17}
{Ferland} G.~J.,  et~al., 2017, \rmxaa, \href
  {http://adsabs.harvard.edu/abs/2017RMxAA..53..385F} {53, 385}

\bibitem[\protect\citeauthoryear{{Finlator}, {Dav{\'e}}  \&
  {Oppenheimer}}{{Finlator} et~al.}{2007}]{finlator+07}
{Finlator} K.,  {Dav{\'e}} R.,   {Oppenheimer} B.~D.,  2007, \mn@doi [\mnras]
  {10.1111/j.1365-2966.2007.11578.x}, \href
  {http://adsabs.harvard.edu/abs/2007MNRAS.376.1861F} {376, 1861}

\bibitem[\protect\citeauthoryear{{Fumagalli} et~al.,}{{Fumagalli}
  et~al.}{2012}]{fumagalli+12}
{Fumagalli} M.,  et~al., 2012, \mn@doi [\apjl] {10.1088/2041-8205/757/2/L22},
  \href {http://adsabs.harvard.edu/abs/2012ApJ...757L..22F} {757, L22}

\bibitem[\protect\citeauthoryear{{Giavalisco} et~al.,}{{Giavalisco}
  et~al.}{2004}]{giavalisco+04a}
{Giavalisco} M.,  et~al., 2004, \mn@doi [\apjl] {10.1086/381244}, \href
  {http://adsabs.harvard.edu/abs/2004ApJ...600L.103G} {600, L103}

\bibitem[\protect\citeauthoryear{{Gonz{\'a}lez}, {Labb{\'e}}, {Bouwens},
  {Illingworth}, {Franx}, {Kriek}  \& {Brammer}}{{Gonz{\'a}lez}
  et~al.}{2010}]{gonzalezetal2010}
{Gonz{\'a}lez} V.,  {Labb{\'e}} I.,  {Bouwens} R.~J.,  {Illingworth} G.,
  {Franx} M.,  {Kriek} M.,   {Brammer} G.~B.,  2010, \mn@doi [\apj]
  {10.1088/0004-637X/713/1/115}, \href
  {http://adsabs.harvard.edu/abs/2010ApJ...713..115G} {713, 115}

\bibitem[\protect\citeauthoryear{{Gonz{\'a}lez}, {Labb{\'e}}, {Bouwens},
  {Illingworth}, {Franx}  \& {Kriek}}{{Gonz{\'a}lez}
  et~al.}{2011}]{gonzalez+11}
{Gonz{\'a}lez} V.,  {Labb{\'e}} I.,  {Bouwens} R.~J.,  {Illingworth} G.,
  {Franx} M.,   {Kriek} M.,  2011, \mn@doi [\apjl]
  {10.1088/2041-8205/735/2/L34}, \href
  {http://adsabs.harvard.edu/abs/2011ApJ...735L..34G} {735, L34+}

\bibitem[\protect\citeauthoryear{{Grazian} et~al.,}{{Grazian}
  et~al.}{2015}]{grazian+15}
{Grazian} A.,  et~al., 2015, \mn@doi [\aap] {10.1051/0004-6361/201424750},
  \href {http://adsabs.harvard.edu/abs/2015A%26A...575A..96G} {575, A96}

\bibitem[\protect\citeauthoryear{{Grogin} et~al.,}{{Grogin}
  et~al.}{2011}]{grogin+11}
{Grogin} N.~A.,  et~al., 2011, \mn@doi [\apjs] {10.1088/0067-0049/197/2/35},
  \href {http://adsabs.harvard.edu/abs/2011ApJS..197...35G} {197, 35}

\bibitem[\protect\citeauthoryear{{Gutkin}, {Charlot}  \& {Bruzual}}{{Gutkin}
  et~al.}{2016}]{gutkin+16}
{Gutkin} J.,  {Charlot} S.,   {Bruzual} G.,  2016, \mn@doi [\mnras]
  {10.1093/mnras/stw1716}, \href
  {http://adsabs.harvard.edu/abs/2016MNRAS.462.1757G} {462, 1757}

\bibitem[\protect\citeauthoryear{{Hippelein} et~al.,}{{Hippelein}
  et~al.}{2003}]{hippelein+03}
{Hippelein} H.,  et~al., 2003, \mn@doi [\aap] {10.1051/0004-6361:20021898},
  \href {http://adsabs.harvard.edu/abs/2003A%26A...402...65H} {402, 65}

\bibitem[\protect\citeauthoryear{{Illingworth} et~al.,}{{Illingworth}
  et~al.}{2013}]{illingworth+13}
{Illingworth} G.~D.,  et~al., 2013, \mn@doi [\apjs]
  {10.1088/0067-0049/209/1/6}, \href
  {http://adsabs.harvard.edu/abs/2013ApJS..209....6I} {209, 6}

\bibitem[\protect\citeauthoryear{{Inami} et~al.,}{{Inami}
  et~al.}{2017}]{inami+17}
{Inami} H.,  et~al., 2017, \mn@doi [\aap] {10.1051/0004-6361/201731195}, \href
  {http://adsabs.harvard.edu/abs/2017A%26A...608A...2I} {608, A2}

\bibitem[\protect\citeauthoryear{{Izotov}, {Guseva}  \& {Thuan}}{{Izotov}
  et~al.}{2011}]{izotov+11}
{Izotov} Y.~I.,  {Guseva} N.~G.,   {Thuan} T.~X.,  2011, \mn@doi [\apj]
  {10.1088/0004-637X/728/2/161}, \href
  {http://adsabs.harvard.edu/abs/2011ApJ...728..161I} {728, 161}

\bibitem[\protect\citeauthoryear{{Izotov}, {Guseva}, {Fricke}, {Henkel}  \&
  {Schaerer}}{{Izotov} et~al.}{2017}]{izotov+17}
{Izotov} Y.~I.,  {Guseva} N.~G.,  {Fricke} K.~J.,  {Henkel} C.,   {Schaerer}
  D.,  2017, \mn@doi [\mnras] {10.1093/mnras/stx347}, \href
  {http://adsabs.harvard.edu/abs/2017MNRAS.467.4118I} {467, 4118}

\bibitem[\protect\citeauthoryear{{Jaskot} \& {Ravindranath}}{{Jaskot} \&
  {Ravindranath}}{2016}]{jaskotravindranath16}
{Jaskot} A.~E.,  {Ravindranath} S.,  2016, \mn@doi [\apj]
  {10.3847/1538-4357/833/2/136}, \href
  {http://adsabs.harvard.edu/abs/2016ApJ...833..136J} {833, 136}

\bibitem[\protect\citeauthoryear{{Kauffmann}, {White}, {Heckman}, {M{\'e}nard},
  {Brinchmann}, {Charlot}, {Tremonti}  \& {Brinkmann}}{{Kauffmann}
  et~al.}{2004}]{kauffmann+04}
{Kauffmann} G.,  {White} S.~D.~M.,  {Heckman} T.~M.,  {M{\'e}nard} B.,
  {Brinchmann} J.,  {Charlot} S.,  {Tremonti} C.,   {Brinkmann} J.,  2004,
  \mn@doi [\mnras] {10.1111/j.1365-2966.2004.08117.x}, \href
  {http://adsabs.harvard.edu/abs/2004MNRAS.353..713K} {353, 713}

\bibitem[\protect\citeauthoryear{{Kewley} \& {Dopita}}{{Kewley} \&
  {Dopita}}{2002}]{kewleydopita02}
{Kewley} L.~J.,  {Dopita} M.~A.,  2002, \mn@doi [\apjs] {10.1086/341326}, \href
  {http://adsabs.harvard.edu/abs/2002ApJS..142...35K} {142, 35}

\bibitem[\protect\citeauthoryear{{Khostovan}, {Sobral}, {Mobasher}, {Best},
  {Smail}, {Stott}, {Hemmati}  \& {Nayyeri}}{{Khostovan}
  et~al.}{2015}]{khostovan+15}
{Khostovan} A.~A.,  {Sobral} D.,  {Mobasher} B.,  {Best} P.~N.,  {Smail} I.,
  {Stott} J.~P.,  {Hemmati} S.,   {Nayyeri} H.,  2015, \mn@doi [\mnras]
  {10.1093/mnras/stv1474}, \href
  {http://adsabs.harvard.edu/abs/2015MNRAS.452.3948K} {452, 3948}

\bibitem[\protect\citeauthoryear{{Khostovan}, {Sobral}, {Mobasher}, {Smail},
  {Darvish}, {Nayyeri}, {Hemmati}  \& {Stott}}{{Khostovan}
  et~al.}{2016}]{khostovan+16}
{Khostovan} A.~A.,  {Sobral} D.,  {Mobasher} B.,  {Smail} I.,  {Darvish} B.,
  {Nayyeri} H.,  {Hemmati} S.,   {Stott} J.~P.,  2016, \mn@doi [\mnras]
  {10.1093/mnras/stw2174}, \href
  {http://adsabs.harvard.edu/abs/2016MNRAS.463.2363K} {463, 2363}

\bibitem[\protect\citeauthoryear{{Koekemoer} et~al.,}{{Koekemoer}
  et~al.}{2011}]{koekemoer+11}
{Koekemoer} A.~M.,  et~al., 2011, \mn@doi [\apjs] {10.1088/0067-0049/197/2/36},
  \href {http://adsabs.harvard.edu/abs/2011ApJS..197...36K} {197, 36}

\bibitem[\protect\citeauthoryear{{Labb{\'e}} et~al.,}{{Labb{\'e}}
  et~al.}{2010}]{labbe+10}
{Labb{\'e}} I.,  et~al., 2010, \mn@doi [\apjl] {10.1088/2041-8205/716/2/L103},
  \href {http://adsabs.harvard.edu/abs/2010ApJ...716L.103L} {716, L103}

\bibitem[\protect\citeauthoryear{{Labb{\'e}} et~al.,}{{Labb{\'e}}
  et~al.}{2013}]{labbe+13}
{Labb{\'e}} I.,  et~al., 2013, \mn@doi [\apjl] {10.1088/2041-8205/777/2/L19},
  \href {http://adsabs.harvard.edu/abs/2013ApJ...777L..19L} {777, L19}

\bibitem[\protect\citeauthoryear{{Labb{\'e}} et~al.,}{{Labb{\'e}}
  et~al.}{2015}]{labbe+15}
{Labb{\'e}} I.,  et~al., 2015, \mn@doi [\apjs] {10.1088/0067-0049/221/2/23},
  \href {http://adsabs.harvard.edu/abs/2015ApJS..221...23L} {221, 23}

\bibitem[\protect\citeauthoryear{{Lam} et~al.,}{{Lam} et~al.}{2019}]{lam+19}
{Lam} D.,  et~al., 2019, arXiv e-prints, \href
  {http://adsabs.harvard.edu/abs/2019arXiv190202786L} {}

\bibitem[\protect\citeauthoryear{{Ly} et~al.,}{{Ly} et~al.}{2007}]{ly+07}
{Ly} C.,  et~al., 2007, \mn@doi [\apj] {10.1086/510828}, \href
  {http://adsabs.harvard.edu/abs/2007ApJ...657..738L} {657, 738}

\bibitem[\protect\citeauthoryear{{Madau} \& {Dickinson}}{{Madau} \&
  {Dickinson}}{2014}]{madaudickinson14}
{Madau} P.,  {Dickinson} M.,  2014, \mn@doi [\araa]
  {10.1146/annurev-astro-081811-125615}, \href
  {http://adsabs.harvard.edu/abs/2014ARA%26A..52..415M} {52, 415}

\bibitem[\protect\citeauthoryear{{M{\'a}rmol-Queralt{\'o}}, {McLure}, {Cullen},
  {Dunlop}, {Fontana}  \& {McLeod}}{{M{\'a}rmol-Queralt{\'o}}
  et~al.}{2016}]{marmol+16}
{M{\'a}rmol-Queralt{\'o}} E.,  {McLure} R.~J.,  {Cullen} F.,  {Dunlop} J.~S.,
  {Fontana} A.,   {McLeod} D.~J.,  2016, \mn@doi [\mnras]
  {10.1093/mnras/stw1212}, \href
  {http://adsabs.harvard.edu/abs/2016MNRAS.460.3587M} {460, 3587}

\bibitem[\protect\citeauthoryear{{Mashian} et~al.,}{{Mashian}
  et~al.}{2015}]{mashian+15}
{Mashian} N.,  et~al., 2015, \mn@doi [\apj] {10.1088/0004-637X/802/2/81}, \href
  {http://adsabs.harvard.edu/abs/2015ApJ...802...81M} {802, 81}

\bibitem[\protect\citeauthoryear{{Matthee}, {Sobral}, {Best}, {Khostovan},
  {Oteo}, {Bouwens}  \& {R{\"o}ttgering}}{{Matthee}
  et~al.}{2017a}]{matthee+17b}
{Matthee} J.,  {Sobral} D.,  {Best} P.,  {Khostovan} A.~A.,  {Oteo} I.,
  {Bouwens} R.,   {R{\"o}ttgering} H.,  2017a, \mn@doi [\mnras]
  {10.1093/mnras/stw2973}, \href
  {http://adsabs.harvard.edu/abs/2017MNRAS.465.3637M} {465, 3637}

\bibitem[\protect\citeauthoryear{{Matthee}, {Sobral}, {Darvish}, {Santos},
  {Mobasher}, {Paulino-Afonso}, {R{\"o}ttgering}  \& {Alegre}}{{Matthee}
  et~al.}{2017b}]{matthee+17}
{Matthee} J.,  {Sobral} D.,  {Darvish} B.,  {Santos} S.,  {Mobasher} B.,
  {Paulino-Afonso} A.,  {R{\"o}ttgering} H.,   {Alegre} L.,  2017b, \mn@doi
  [\mnras] {10.1093/mnras/stx2061}, \href
  {http://adsabs.harvard.edu/abs/2017MNRAS.472..772M} {472, 772}

\bibitem[\protect\citeauthoryear{{Nakajima} et~al.,}{{Nakajima}
  et~al.}{2018}]{nakajima+18}
{Nakajima} K.,  et~al., 2018, \mn@doi [\aap] {10.1051/0004-6361/201731935},
  \href {http://adsabs.harvard.edu/abs/2018A%26A...612A..94N} {612, A94}

\bibitem[\protect\citeauthoryear{{Oesch} et~al.,}{{Oesch}
  et~al.}{2015}]{oesch+15}
{Oesch} P.~A.,  et~al., 2015, \mn@doi [\apjl] {10.1088/2041-8205/804/2/L30},
  \href {http://adsabs.harvard.edu/abs/2015ApJ...804L..30O} {804, L30}

\bibitem[\protect\citeauthoryear{{Oke} \& {Gunn}}{{Oke} \&
  {Gunn}}{1983}]{okegunn83}
{Oke} J.~B.,  {Gunn} J.~E.,  1983, \mn@doi [\apj] {10.1086/160817}, \href
  {http://adsabs.harvard.edu/abs/1983ApJ...266..713O} {266, 713}

\bibitem[\protect\citeauthoryear{{Pirzkal} et~al.,}{{Pirzkal}
  et~al.}{2013}]{pirzkal+13}
{Pirzkal} N.,  et~al., 2013, \mn@doi [\apj] {10.1088/0004-637X/772/1/48}, \href
  {http://adsabs.harvard.edu/abs/2013ApJ...772...48P} {772, 48}

\bibitem[\protect\citeauthoryear{{Prevot}, {Lequeux}, {Prevot}, {Maurice}  \&
  {Rocca-Volmerange}}{{Prevot} et~al.}{1984}]{prevot+84}
{Prevot} M.~L.,  {Lequeux} J.,  {Prevot} L.,  {Maurice} E.,
  {Rocca-Volmerange} B.,  1984, \aap, \href
  {http://adsabs.harvard.edu/abs/1984A%26A...132..389P} {132, 389}

\bibitem[\protect\citeauthoryear{{Rasappu}, {Smit}, {Labb{\'e}}, {Bouwens},
  {Stark}, {Ellis}  \& {Oesch}}{{Rasappu} et~al.}{2016}]{rasappu+16}
{Rasappu} N.,  {Smit} R.,  {Labb{\'e}} I.,  {Bouwens} R.~J.,  {Stark} D.~P.,
  {Ellis} R.~S.,   {Oesch} P.~A.,  2016, \mn@doi [\mnras]
  {10.1093/mnras/stw1484}, \href
  {http://adsabs.harvard.edu/abs/2016MNRAS.461.3886R} {461, 3886}

\bibitem[\protect\citeauthoryear{{Reddy} \& {Steidel}}{{Reddy} \&
  {Steidel}}{2009}]{reddysteidel09}
{Reddy} N.~A.,  {Steidel} C.~C.,  2009, \mn@doi [\apj]
  {10.1088/0004-637X/692/1/778}, \href
  {http://adsabs.harvard.edu/abs/2009ApJ...692..778R} {692, 778}

\bibitem[\protect\citeauthoryear{{Reddy} et~al.,}{{Reddy}
  et~al.}{2015}]{reddy+15}
{Reddy} N.~A.,  et~al., 2015, \mn@doi [\apj] {10.1088/0004-637X/806/2/259},
  \href {http://adsabs.harvard.edu/abs/2015ApJ...806..259R} {806, 259}

\bibitem[\protect\citeauthoryear{{Rigby}, {Bayliss}, {Gladders}, {Sharon},
  {Wuyts}, {Dahle}, {Johnson}  \& {Pe{\~n}a-Guerrero}}{{Rigby}
  et~al.}{2015}]{rigby+15}
{Rigby} J.~R.,  {Bayliss} M.~B.,  {Gladders} M.~D.,  {Sharon} K.,  {Wuyts} E.,
  {Dahle} H.,  {Johnson} T.,   {Pe{\~n}a-Guerrero} M.,  2015, \mn@doi [\apjl]
  {10.1088/2041-8205/814/1/L6}, \href
  {http://adsabs.harvard.edu/abs/2015ApJ...814L...6R} {814, L6}

\bibitem[\protect\citeauthoryear{{Roberts-Borsani} et~al.,}{{Roberts-Borsani}
  et~al.}{2016}]{robertsborsani+16}
{Roberts-Borsani} G.~W.,  et~al., 2016, \mn@doi [\apj]
  {10.3847/0004-637X/823/2/143}, \href
  {http://adsabs.harvard.edu/abs/2016ApJ...823..143R} {823, 143}

\bibitem[\protect\citeauthoryear{{Salmon} et~al.,}{{Salmon}
  et~al.}{2015}]{salmon+15}
{Salmon} B.,  et~al., 2015, \mn@doi [\apj] {10.1088/0004-637X/799/2/183}, \href
  {http://adsabs.harvard.edu/abs/2015ApJ...799..183S} {799, 183}

\bibitem[\protect\citeauthoryear{{Salvatier}, {Wiecki}  \&
  {Fonnesbeck}}{{Salvatier} et~al.}{2016}]{pymc3}
{Salvatier} J.,  {Wiecki} T.~V.,   {Fonnesbeck} C.,  2016, {PyMC3: Python
  probabilistic programming framework}, Astrophysics Source Code Library
  (\mn@eprint {ascl} {1610.016})

\bibitem[\protect\citeauthoryear{{Schaerer} \& {De Barros}}{{Schaerer} \& {De
  Barros}}{2009}]{schaererdebarros09}
{Schaerer} D.,  {De Barros} S.,  2009, \mn@doi [\aap]
  {10.1051/0004-6361/200911781}, \href
  {http://adsabs.harvard.edu/abs/2009A%26A...502..423S} {502, 423}

\bibitem[\protect\citeauthoryear{{Schaerer} \& {De Barros}}{{Schaerer} \& {De
  Barros}}{2010}]{schaererdebarros10}
{Schaerer} D.,  {De Barros} S.,  2010, \mn@doi [\aap]
  {10.1051/0004-6361/200913946}, \href
  {http://adsabs.harvard.edu/abs/2010A%26A...515A..73S} {515, A73+}

\bibitem[\protect\citeauthoryear{{Schaerer}, {Izotov}, {Verhamme},
  {Orlitov{\'a}}, {Thuan}, {Worseck}  \& {Guseva}}{{Schaerer}
  et~al.}{2016}]{schaerer+16}
{Schaerer} D.,  {Izotov} Y.~I.,  {Verhamme} A.,  {Orlitov{\'a}} I.,  {Thuan}
  T.~X.,  {Worseck} G.,   {Guseva} N.~G.,  2016, \mn@doi [\aap]
  {10.1051/0004-6361/201628943}, \href
  {http://adsabs.harvard.edu/abs/2016A%26A...591L...8S} {591, L8}

\bibitem[\protect\citeauthoryear{{Senchyna} et~al.,}{{Senchyna}
  et~al.}{2017}]{senchyna+17}
{Senchyna} P.,  et~al., 2017, \mn@doi [\mnras] {10.1093/mnras/stx2059}, \href
  {http://adsabs.harvard.edu/abs/2017MNRAS.472.2608S} {472, 2608}

\bibitem[\protect\citeauthoryear{{Sharma}}{{Sharma}}{2017}]{sharma17}
{Sharma} S.,  2017, \mn@doi [\araa] {10.1146/annurev-astro-082214-122339},
  \href {http://adsabs.harvard.edu/abs/2017ARA%26A..55..213S} {55, 213}

\bibitem[\protect\citeauthoryear{{Shim}, {Chary}, {Dickinson}, {Lin},
  {Spinrad}, {Stern}  \& {Yan}}{{Shim} et~al.}{2011}]{shim+11}
{Shim} H.,  {Chary} R.-R.,  {Dickinson} M.,  {Lin} L.,  {Spinrad} H.,  {Stern}
  D.,   {Yan} C.-H.,  2011, \mn@doi [\apj] {10.1088/0004-637X/738/1/69}, \href
  {http://adsabs.harvard.edu/abs/2011ApJ...738...69S} {738, 69}

\bibitem[\protect\citeauthoryear{{Shivaei}, {Reddy}, {Steidel}  \&
  {Shapley}}{{Shivaei} et~al.}{2015}]{shivaei+15a}
{Shivaei} I.,  {Reddy} N.~A.,  {Steidel} C.~C.,   {Shapley} A.~E.,  2015,
  \mn@doi [\apj] {10.1088/0004-637X/804/2/149}, \href
  {http://adsabs.harvard.edu/abs/2015ApJ...804..149S} {804, 149}

\bibitem[\protect\citeauthoryear{{Shivaei} et~al.,}{{Shivaei}
  et~al.}{2018}]{shivaei+18}
{Shivaei} I.,  et~al., 2018, \mn@doi [\apj] {10.3847/1538-4357/aaad62}, \href
  {http://adsabs.harvard.edu/abs/2018ApJ...855...42S} {855, 42}

\bibitem[\protect\citeauthoryear{{Smit}, {Bouwens}, {Franx}, {Illingworth},
  {Labb{\'e}}, {Oesch}  \& {van Dokkum}}{{Smit} et~al.}{2012}]{smit+12}
{Smit} R.,  {Bouwens} R.~J.,  {Franx} M.,  {Illingworth} G.~D.,  {Labb{\'e}}
  I.,  {Oesch} P.~A.,   {van Dokkum} P.~G.,  2012, \mn@doi [\apj]
  {10.1088/0004-637X/756/1/14}, \href
  {http://adsabs.harvard.edu/abs/2012ApJ...756...14S} {756, 14}

\bibitem[\protect\citeauthoryear{{Smit} et~al.,}{{Smit} et~al.}{2014}]{smit+14}
{Smit} R.,  et~al., 2014, \mn@doi [\apj] {10.1088/0004-637X/784/1/58}, \href
  {http://adsabs.harvard.edu/abs/2014ApJ...784...58S} {784, 58}

\bibitem[\protect\citeauthoryear{{Smit} et~al.,}{{Smit} et~al.}{2015}]{smit+15}
{Smit} R.,  et~al., 2015, \mn@doi [\apj] {10.1088/0004-637X/801/2/122}, \href
  {http://adsabs.harvard.edu/abs/2015ApJ...801..122S} {801, 122}

\bibitem[\protect\citeauthoryear{{Smit}, {Bouwens}, {Labb{\'e}}, {Franx},
  {Wilkins}  \& {Oesch}}{{Smit} et~al.}{2016}]{smit+16}
{Smit} R.,  {Bouwens} R.~J.,  {Labb{\'e}} I.,  {Franx} M.,  {Wilkins} S.~M.,
  {Oesch} P.~A.,  2016, \mn@doi [\apj] {10.3847/1538-4357/833/2/254}, \href
  {http://adsabs.harvard.edu/abs/2016ApJ...833..254S} {833, 254}

\bibitem[\protect\citeauthoryear{{Smit}, {Swinbank}, {Massey}, {Richard},
  {Smail}  \& {Kneib}}{{Smit} et~al.}{2017}]{smit+17}
{Smit} R.,  {Swinbank} A.~M.,  {Massey} R.,  {Richard} J.,  {Smail} I.,
  {Kneib} J.-P.,  2017, \mn@doi [\mnras] {10.1093/mnras/stx245}, \href
  {http://adsabs.harvard.edu/abs/2017MNRAS.467.3306S} {467, 3306}

\bibitem[\protect\citeauthoryear{{Sobral}, {Best}, {Smail}, {Mobasher}, {Stott}
   \& {Nisbet}}{{Sobral} et~al.}{2014}]{sobral+14}
{Sobral} D.,  {Best} P.~N.,  {Smail} I.,  {Mobasher} B.,  {Stott} J.,
  {Nisbet} D.,  2014, \mn@doi [\mnras] {10.1093/mnras/stt2159}, \href
  {http://adsabs.harvard.edu/abs/2014MNRAS.437.3516S} {437, 3516}

\bibitem[\protect\citeauthoryear{{Sobral} et~al.,}{{Sobral}
  et~al.}{2018}]{sobral+18}
{Sobral} D.,  et~al., 2018, \mn@doi [\mnras] {10.1093/mnras/sty782}, \href
  {http://adsabs.harvard.edu/abs/2018MNRAS.477.2817S} {477, 2817}

\bibitem[\protect\citeauthoryear{{Song}, {Finkelstein}, {Livermore}, {Capak},
  {Dickinson}  \& {Fontana}}{{Song} et~al.}{2016}]{song+16}
{Song} M.,  {Finkelstein} S.~L.,  {Livermore} R.~C.,  {Capak} P.~L.,
  {Dickinson} M.,   {Fontana} A.,  2016, \mn@doi [\apj]
  {10.3847/0004-637X/826/2/113}, \href
  {http://adsabs.harvard.edu/abs/2016ApJ...826..113S} {826, 113}

\bibitem[\protect\citeauthoryear{{Stanway}, {Eldridge}  \& {Becker}}{{Stanway}
  et~al.}{2016}]{stanway+16}
{Stanway} E.~R.,  {Eldridge} J.~J.,   {Becker} G.~D.,  2016, \mn@doi [\mnras]
  {10.1093/mnras/stv2661}, \href
  {http://adsabs.harvard.edu/abs/2016MNRAS.456..485S} {456, 485}

\bibitem[\protect\citeauthoryear{{Stark}, {Ellis}, {Bunker}, {Bundy},
  {Targett}, {Benson}  \& {Lacy}}{{Stark} et~al.}{2009}]{stark+09}
{Stark} D.~P.,  {Ellis} R.~S.,  {Bunker} A.,  {Bundy} K.,  {Targett} T.,
  {Benson} A.,   {Lacy} M.,  2009, \mn@doi [\apj]
  {10.1088/0004-637X/697/2/1493}, \href
  {http://adsabs.harvard.edu/abs/2009ApJ...697.1493S} {697, 1493}

\bibitem[\protect\citeauthoryear{{Stark}, {Schenker}, {Ellis}, {Robertson},
  {McLure}  \& {Dunlop}}{{Stark} et~al.}{2013}]{stark+13}
{Stark} D.~P.,  {Schenker} M.~A.,  {Ellis} R.,  {Robertson} B.,  {McLure} R.,
  {Dunlop} J.,  2013, \mn@doi [\apj] {10.1088/0004-637X/763/2/129}, \href
  {http://adsabs.harvard.edu/abs/2013ApJ...763..129S} {763, 129}

\bibitem[\protect\citeauthoryear{{Stark} et~al.,}{{Stark}
  et~al.}{2014}]{stark+14}
{Stark} D.~P.,  et~al., 2014, \mn@doi [\mnras] {10.1093/mnras/stu1618}, \href
  {http://adsabs.harvard.edu/abs/2014MNRAS.445.3200S} {445, 3200}

\bibitem[\protect\citeauthoryear{{Stark} et~al.,}{{Stark}
  et~al.}{2015a}]{stark+15a}
{Stark} D.~P.,  et~al., 2015a, \mn@doi [\mnras] {10.1093/mnras/stv688}, \href
  {http://adsabs.harvard.edu/abs/2015MNRAS.450.1846S} {450, 1846}

\bibitem[\protect\citeauthoryear{{Stark} et~al.,}{{Stark}
  et~al.}{2015b}]{stark+15b}
{Stark} D.~P.,  et~al., 2015b, \mn@doi [\mnras] {10.1093/mnras/stv1907}, \href
  {http://adsabs.harvard.edu/abs/2015MNRAS.454.1393S} {454, 1393}

\bibitem[\protect\citeauthoryear{{Stark} et~al.,}{{Stark}
  et~al.}{2017}]{stark+17}
{Stark} D.~P.,  et~al., 2017, \mn@doi [\mnras] {10.1093/mnras/stw2233}, \href
  {http://adsabs.harvard.edu/abs/2017MNRAS.464..469S} {464, 469}

\bibitem[\protect\citeauthoryear{{Stefanon}, {Bouwens}, {Labb{\'e}}, {Muzzin},
  {Marchesini}, {Oesch}  \& {Gonzalez}}{{Stefanon} et~al.}{2017}]{stefanon+17}
{Stefanon} M.,  {Bouwens} R.~J.,  {Labb{\'e}} I.,  {Muzzin} A.,  {Marchesini}
  D.,  {Oesch} P.,   {Gonzalez} V.,  2017, \mn@doi [\apj]
  {10.3847/1538-4357/aa72d8}, \href
  {http://adsabs.harvard.edu/abs/2017ApJ...843...36S} {843, 36}

\bibitem[\protect\citeauthoryear{{Steidel}, {Giavalisco}, {Pettini},
  {Dickinson}  \& {Adelberger}}{{Steidel} et~al.}{1996}]{steidel+96}
{Steidel} C.~C.,  {Giavalisco} M.,  {Pettini} M.,  {Dickinson} M.,
  {Adelberger} K.~L.,  1996, \mn@doi [\apjl] {10.1086/310029}, \href
  {http://adsabs.harvard.edu/abs/1996ApJ...462L..17S} {462, L17}

\bibitem[\protect\citeauthoryear{{Steidel}, {Strom}, {Pettini}, {Rudie},
  {Reddy}  \& {Trainor}}{{Steidel} et~al.}{2016}]{steidel+16}
{Steidel} C.~C.,  {Strom} A.~L.,  {Pettini} M.,  {Rudie} G.~C.,  {Reddy} N.~A.,
    {Trainor} R.~F.,  2016, \mn@doi [\apj] {10.3847/0004-637X/826/2/159}, \href
  {http://adsabs.harvard.edu/abs/2016ApJ...826..159S} {826, 159}

\bibitem[\protect\citeauthoryear{{Storey} \& {Hummer}}{{Storey} \&
  {Hummer}}{1995}]{storeyhummer95}
{Storey} P.~J.,  {Hummer} D.~G.,  1995, \mnras, \href
  {http://adsabs.harvard.edu/abs/1995MNRAS.272...41S} {272, 41}

\bibitem[\protect\citeauthoryear{{Tang}, {Stark}, {Chevallard}  \&
  {Charlot}}{{Tang} et~al.}{2018}]{tang+18}
{Tang} M.,  {Stark} D.,  {Chevallard} J.,   {Charlot} S.,  2018, preprint,
  \href {http://adsabs.harvard.edu/abs/2018arXiv180909637T} {} (\mn@eprint
  {arXiv} {1809.09637})

\bibitem[\protect\citeauthoryear{{The Astropy Collaboration} et~al.,}{{The
  Astropy Collaboration} et~al.}{2018}]{astropy}
{The Astropy Collaboration} et~al., 2018, \mn@doi [\aj]
  {10.3847/1538-3881/aabc4f}, \href
  {http://adsabs.harvard.edu/abs/2018AJ....156..123T} {156, 123}

\bibitem[\protect\citeauthoryear{{Theios}, {Steidel}, {Strom}, {Rudie},
  {Trainor}  \& {Reddy}}{{Theios} et~al.}{2019}]{theios+19}
{Theios} R.~L.,  {Steidel} C.~C.,  {Strom} A.~L.,  {Rudie} G.~C.,  {Trainor}
  R.~F.,   {Reddy} N.~A.,  2019, \mn@doi [\apj] {10.3847/1538-4357/aaf386},
  \href {http://adsabs.harvard.edu/abs/2019ApJ...871..128T} {871, 128}

\bibitem[\protect\citeauthoryear{{Tremonti} et~al.,}{{Tremonti}
  et~al.}{2004}]{tremonti+04}
{Tremonti} C.~A.,  et~al., 2004, \mn@doi [\apj] {10.1086/423264}, \href
  {http://adsabs.harvard.edu/abs/2004ApJ...613..898T} {613, 898}

\bibitem[\protect\citeauthoryear{{Vanzella} et~al.,}{{Vanzella}
  et~al.}{2017}]{vanzella+17}
{Vanzella} E.,  et~al., 2017, \mn@doi [\apj] {10.3847/1538-4357/aa74ae}, \href
  {http://adsabs.harvard.edu/abs/2017ApJ...842...47V} {842, 47}

\bibitem[\protect\citeauthoryear{{Williams} et~al.,}{{Williams}
  et~al.}{2018}]{williams+18}
{Williams} C.~C.,  et~al., 2018, \mn@doi [\apjs] {10.3847/1538-4365/aabcbb},
  \href {http://adsabs.harvard.edu/abs/2018ApJS..236...33W} {236, 33}

\bibitem[\protect\citeauthoryear{{Wofford} et~al.,}{{Wofford}
  et~al.}{2016}]{wofford+16}
{Wofford} A.,  et~al., 2016, \mn@doi [\mnras] {10.1093/mnras/stw150}, \href
  {http://adsabs.harvard.edu/abs/2016MNRAS.457.4296W} {457, 4296}

\bibitem[\protect\citeauthoryear{{Yabe}, {Ohta}, {Iwata}, {Sawicki}, {Tamura},
  {Akiyama}  \& {Aoki}}{{Yabe} et~al.}{2009}]{yabe+09}
{Yabe} K.,  {Ohta} K.,  {Iwata} I.,  {Sawicki} M.,  {Tamura} N.,  {Akiyama} M.,
    {Aoki} K.,  2009, \mn@doi [\apj] {10.1088/0004-637X/693/1/507}, \href
  {http://adsabs.harvard.edu/abs/2009ApJ...693..507Y} {693, 507}

\bibitem[\protect\citeauthoryear{{Yang}, {Malhotra}, {Rhoads}  \&
  {Wang}}{{Yang} et~al.}{2017}]{yang+17b}
{Yang} H.,  {Malhotra} S.,  {Rhoads} J.~E.,   {Wang} J.,  2017, \mn@doi [\apj]
  {10.3847/1538-4357/aa8809}, \href
  {http://adsabs.harvard.edu/abs/2017ApJ...847...38Y} {847, 38}

\bibitem[\protect\citeauthoryear{{Zackrisson}, {Bergvall}, {Olofsson}  \&
  {Siebert}}{{Zackrisson} et~al.}{2001}]{zackrisson+01}
{Zackrisson} E.,  {Bergvall} N.,  {Olofsson} K.,   {Siebert} A.,  2001, \mn@doi
  [\aap] {10.1051/0004-6361:20010912}, \href
  {http://adsabs.harvard.edu/abs/2001A%26A...375..814Z} {375, 814}

\bibitem[\protect\citeauthoryear{{Zackrisson}, {Rydberg}, {Schaerer},
  {{\"O}stlin}  \& {Tuli}}{{Zackrisson} et~al.}{2011}]{zackrisson+11}
{Zackrisson} E.,  {Rydberg} C.-E.,  {Schaerer} D.,  {{\"O}stlin} G.,   {Tuli}
  M.,  2011, \mn@doi [\apj] {10.1088/0004-637X/740/1/13}, \href
  {http://adsabs.harvard.edu/abs/2011ApJ...740...13Z} {740, 13}

\bibitem[\protect\citeauthoryear{{Zitrin} et~al.,}{{Zitrin}
  et~al.}{2015}]{zitrin+15}
{Zitrin} A.,  et~al., 2015, \mn@doi [\apjl] {10.1088/2041-8205/810/1/L12},
  \href {http://adsabs.harvard.edu/abs/2015ApJ...810L..12Z} {810, L12}

\makeatother
\end{thebibliography}








\bsp	
\label{lastpage}
\end{document}